\documentclass[onecolumn,amsmath,amssymb,12pt]{article}

\usepackage{amsmath, amsfonts}
\usepackage{graphicx}
\usepackage{color}
\usepackage{indentfirst}
\usepackage{float}
\usepackage{graphicx}
\usepackage{multirow}
\usepackage[dvipsnames]{xcolor}
\usepackage[a4paper]{geometry}
\usepackage[footnotesize,bf]{caption2}
\usepackage{footmisc}
\newcommand{\GB}{\textcolor{black}} 
\newcommand{\AB}{\textcolor{black}} 
\newcommand{\AP}{\textcolor{black}} 
 
\begin{document}
\author{\AB {
	A.E.\ Biondo\thanks{Dept.\ of Economics and Business, University of Catania, Italy; corresponding: ae.biondo@unict.it},
	G.\ Burgio\thanks{Dept.\ of Computer Engineering and Mathematics, University Rovira i Virgili, Spain},
	A.\ Pluchino\thanks{Dept.\ of Physics and Astronomy, University of Catania and INFN Sezione di Catania, Italy},
	D.\ Puglisi\thanks{Dept.\ of Physics and Astronomy, University of Catania, Italy}.
	}
}
\title{\AB{A dynamic model of Tax Evasion }}

\date{ }
\maketitle

\begin{abstract}
	In this paper we present a simplified model of a proportional taxation system where citizens decide whether to pay taxes or evade them.  We initially derive a dynamic equation for the fraction of evaders and then present both its critical points and equilibrium stable states for a well-mixed population and for a fixed audit probability. Our theoretical predictions consider different possible income distributions (homogeneous and heterogeneous), time-dependent audit probability and different possible social topologies, by means of selected complex network configurations. All derived results are validated and confirmed by extended Monte Carlo simulations. Finally, some policy implications aimed to reduce tax evasion are suggested, with regards to the tax and fine rates and the audit probability.\\
	
	\textit{JEL} codes: C62, C63, H26. \\
	
	\textit{Keywords}: Tax evasion, Tax Policy, Monte Carlo Simulations, Complex Networks. 
\end{abstract}

\section{Introduction}	
\label{sec:intro}
\AB{Almost all economic systems suffer from the tax evasion, which emerges as a collective problem deriving from the aggregation of selfish behaviors assumed by \textit{free riders}, i.e., citizens consuming public goods without contributing to their costs. The phenomenon and, more broadly, the black economy, is not identical everywhere, being connected to the degree of civic maturity of populations. Related literature has shown that, for example, people are more likely to evade if they perceive that others are evading, especially in presence of not too high probabilities of being caught (see, for example, Alm et al. (2015), Dubin (2007), McGee (2012), Torgler (2007)). Notable differences can be observed also depending on countries, as in Pickhardt and Prinz (2012), Riahi-Belkaoui (2004), and Torgler and Schaltegger (2005), while Kirchler (2007) and Shu et al. (2012) offer a focus on a series of factors affecting tax compliance, such as perceptions about tax fairness, attitudes and trust toward government, social and personal norms. In some cases, tax evasion can lead to very perverse systemic effects, such as severe limitations to the possibilities of governmental expenditure decisions, reduction in the quality and the quantity of supplied services, inefficient and negative redistributive consequences, growth of the shadow economy. Several contributes deal with consequences of the phenomenon, as in Andreoni \textit{et al.} (1998), Slemrod and Yitzhaki (2002), Torgler (2002), Kirchler (2007), Slemrod (2007), among others, while a specific reference to evasion in redistributive terms and the role of income in the decision to evade is in Alstadsaeter \textit{et al.}  (2017), and in Bertotti and Modanese (2014, 2016).}

\AB{Models of tax evasion have been developed since the Seventies, by considering optimizing individuals -- fully informed on audit rates and penalties -- facing different audit strategies. Such contributions discuss also the impact on evasion of different tax rates, income distribution and the choice of the basis for taxes computation. Examples are, among others, Allingham and Sandmo (1972), Yitzhaki (1987), Clotfelter (1983), Crane and Nourzad (1987), Poterba (1987), Panades (2004), Dalamagas (2011).}

\AB{The issue of tax evasion has been approached also through contributions based on agent-based models. A survey of such papers could be gained by the joint reading of Bloomquist (2006), Alm (2012), Hokamp (2013), Pickhardt and Prinz (2014), Bazart \textit{et al}. (2016). The advantage of agent-based models is that they are prone to describe the complexity of aggregate contexts, as documented in previous studies of socio-economic and financial markets analysis, as in Pluchino \textit{et al}. (2010 and 2018), Biondo \textit{et al}. (2013), Squazzoni \textit{et al.} (2014), and Biondo (2019). Simulative models can be particularly helpful in studying behavioral attributes, thus effectively understanding the importance of social norms and auditing, as in Hokamp and Pickhardt (2010), and the effect of social networks on the tax compliance, as in Vale (2015).}

\AB{Policy makers have many reasons to fight tax evasion but, at the same time, politicians cannot assume drastic positions about the topic, specially in contexts where the problem is highly diffuse. The IRS (2016) reports that the expenditures paid by governmental authorities to induce virtuous behaviors are significant. However, in many cases, free riders remain unpunished while providing the bad example of being smart enough to exploit common resources without paying their part. The known conflict between \textit{individual} and \textit{collective} rationality (Rapoport 1974) creates the paradoxical outcome of the \textit{prisoner's dilemma}. A vast amount of literature, regarding the production of public goods, as in Heckathorn (1996), the emergence of social norms and social interaction, as in Hardin (1995), and Voss (2001), among others, pose the question about the individual decision to cooperate. Cooperation could be the optimal choice because of different reasons. First of all, simple altruism, as recalled, for just some examples, in Stevens (2018) and Epstein (1993); secondly, imitation, as in Callen and Shapero (1974), in Elsenbroich and Gilbert (2014) and McDonald and Crandall (2015); alternatively, needed quality of the public good (i.e., quality of Institutions), as in Nicolaides (2014), Feld and Frey (2007) and Torgler and Schneider (2009), among others.}

\AP{The main aim of this paper is to study possible policy directions to reduce tax evasion and provide asymptotic conditions to eradicate it, by relying on a novel proportional fine system, with possibly variable penalties, and on a tunable audit probability. Following an analytical approach applied to a simplified model of proportional taxation/penalization, called Proportional Taxation Model (PTM), we first derive, in Section 2, the dynamic equation which regulates the time evolution of the fraction of tax-evaders, along with its critical points and equilibria. Then, in Section 3, we present results obtained for a well-mixed population and a fixed audit probability, considering both an homogeneous and an heterogeneous income distribution and comparing the theoretical predictions with the outcome of numerical simulations; in the same section, the previous analysis is repeated also by introducing a time-dependent audit probability. Finally, in Section 4, the case of a structured population, with a complex network topology, is addressed, exploring to what extent the social interaction has a room to influence the macro-dynamics of the phenomenon by means of relational feedbacks on the economic status of neighbors. Some final remarks will conclude the paper, in Section 5. Details of analytical methods and calculations are reported in the Appendix.}

\section{Proportional Taxation Model (PTM)}
\label{sec:model}

\AP{Suppose a population of $N$ agents. At time $t$ the generic agent $A_i$, with $i\in\{1,\dots,N\}$, gets a periodic income $c_i(t)$ extracted at the beginning from an income distribution $p(c)$ and fixed for all times, i.e., $c_i(t)=c_i(0)\equiv c_i$, $\forall t$.} \AB{Each agent plays a simplified \textit{public good game} by choosing one of two possible strategies: \textsc{pay} and \textsc{evade}. If \textsc{pay} is chosen, agent $i$'s income, at the end of round $t$, is reduced to $c_i(t)(1-T)$, where $T\in[0,1]$ is the fixed proportional tax; if, instead, strategy \textsc{evade} is chosen, agent $i$'s  final income is $c_i(t)(1-P)$ with probability $p_a$ and by $c_i(t)$ with probability $1-p_a$, where $P\in[0,1]$ is the fixed proportional fine and $p_a\in[0,1]$ is the audit probability.} 

\begin{figure}[t!]
	\centering
		\includegraphics[width=.85\linewidth]{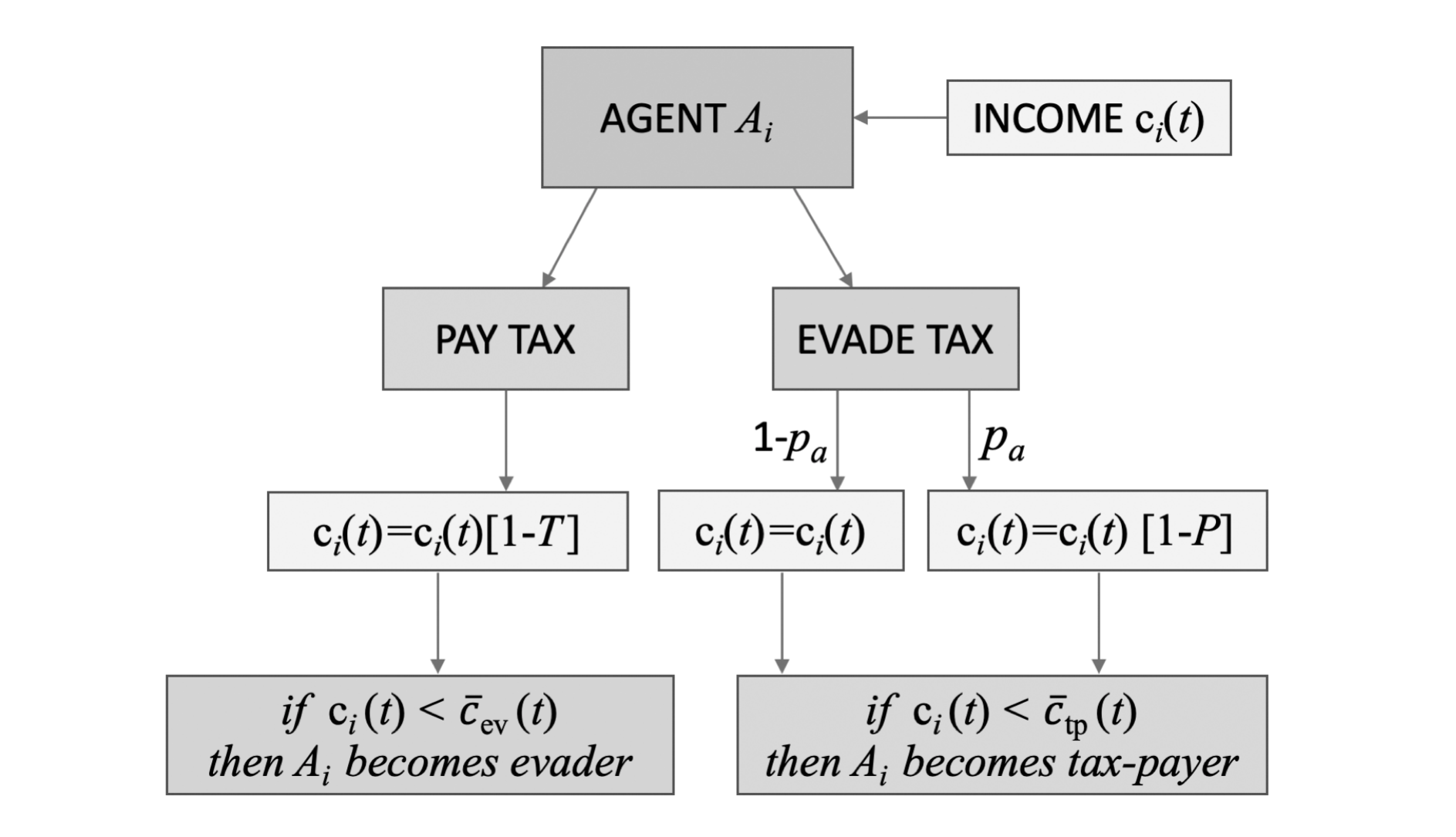}
		\caption{\AB{The single-agent strategy tree of the Proportional Taxation Model (PTM), at time $t$. According to the strategy chosen by agent $A_i$ and the audit probability $p_a$, her periodic income $c_i(t)$ could be reduced of a given amount, proportional to the income itself. After comparing her resulting income with the average one of a given subset of agents playing the opposite strategy (if any), agent $A_i$ will change strategy in $t+\Delta t$ if the former is lower than the latter.}
		}
\label{fig:strategies}
\end{figure}

\AP{Figure~\ref{fig:strategies} depicts the single-agent strategy tree.} \GB{After all agents have played, one of them is randomly chosen and considered to possibly change her strategy for the next time step, $t+\Delta t$. The picked agent interacts with a subset of other agents (as specified below), among which, possibly, tax payers and evaders. Average incomes of both groups are computed, respectively, as $\bar{c}_{tp}(t)$ and $\bar{c}_{ev}(t)$. The agent will re-play the same strategy chosen in $t$ if her current income is greater than or equal to the average income of the group playing the opposite strategy, otherwise she will adopt the latter in $t+\Delta t$.}	
Since exactly one strategy update is considered at each time step, we can take $\Delta t = 1/N$, so that time $t$ parametrizes (as a dimensionless parameter) the evolution of the population by counting how many single updates have been considered until then, relatively to the population size $N$. In order to write down a reliable dynamic equation for the state of the population, we consider the large $N$ limit. In particular, for $N\rightarrow\infty$, $\Delta t\rightarrow 0$ and we get the continuous-time description. Moreover, for the dynamic to be non-trivial, it must hold $P>T$ (i.e., evading taxes must be risky), thus we only consider this case.

The initial strategies and incomes are assigned at random to the players, hence, at least initially, strategy $s$ and income $c$ are uncorrelated variables and we can factorize the joint probability $p(s,c)$ of taking strategy $s$ and getting income $c$ into the product of the marginal probabilities, $p(s)p(c)$. In the following, we assume that factorization to hold for all times. Defining, from now on, $c_-$ and $c_+$ as the minimum and maximum allowed incomes, and with $\rho_c$ the \AP{fraction} of tax-evaders with income $c$, $\rho=\int_{c_-}^{c_+} dc~\rho_c$ is the overall  \AP{fraction} of tax-evaders, and $N\rho(t)$ gives the number of tax-evaders in the population at time $t$. The aforementioned factorization allows to write $\rho_c=\rho~p(c)$. From the normalization constraint it follows $1-\rho$ to be the \AP{overall fraction} of tax-payers and $(1-\rho)p(c)$ the fraction of them receiving income $c$. 

\AB{The scope of this section is to derive a dynamic equation for $\rho(t)$ describing the macroscopic state of the population (subsection \ref{sec:dynamics}), along with the expressions for its critical points and equilibrium stable states (subsection \ref{sec:critical}). To this purpose, we have to define the selection mechanism for the subset of agents used for income comparison with the picked agent. Let us initially consider a population in which any agent can interact with any other agent, i.e., a well-mixed population. The picked agent randomly chooses $ z\in\mathbb{N}^+$ other agents. Note that an agent will generally interact with different sets of other agents at different time steps. In Section~\ref{sec:model_SP}, the case of a population organized in a fixed social structure, governing possible interactions, will be considered}. As it will be shown, the social structure comes out to play a secondary role, hence sublimating the analytically tractable well-mixed population to a reliable approximation, at least when structural heterogeneity among individuals and their incomes are uncorrelated.

\subsection{\AB{Dynamics}}
\label{sec:dynamics}

\AB{In order to derive a dynamic equation for $\rho(t)$, the conditions for strategy changes of any type of agents, i.e., the probability of any configuration of agents with a given set of strategies and average income, must be set. Let us begin with the probability of choosing, randomly, among $z$ agents, $n_e$ tax-evaders, of which $n_e^+$ \lq\lq lucky\rq\rq (i.e., avoiding the audit) and $n_e^-$ \lq\lq unlucky\rq\rq, and $n_p=z-n_e$ tax-payers, regardless of their incomes. It can be computed as
\begin{equation}
	\xi_e \left(z,n_e,n_e^-;p_a\right) = \binom{z}{n_e}\binom{n_e}{n_e^-}\left(1-p_a\right)^{n-n_e^-}p_a^{n_e^-}\left[\rho(t)\right]^{n_e}\left[1-\rho(t)\right]^{z-n_e}
\label{conf_prob_audit}
\end{equation}
Summing over $n_e^-$ from $0$ to $n_e$, we get the probability of interacting with $n_e$ tax-evaders, regardless of the audit outcome, and $n_p$ tax-payers. This can be also written as
\begin{equation}
	\xi_p \left(z,n_p\right) =  \sum_{n_e^- = 0}^{n_e} \xi\left(n_e,n_e^-\right) = \binom{z}{n_p}\left[\rho(t)\right]^{z-n_p}\left[1-\rho(t)\right]^{n_p}
\label{conf_prob}
\end{equation}}

\AB{Consider that, independently of selection time, a group of $n$ selected tax-payers $\{A_1,\dots,A_n\}$, has an average income equal to $$\bar{c}_{tp} = (1-T)(c_{1}+\dots +c_{n})/n$$ while  a group of $n$ tax-evaders has an average income $$\bar{c}_{ev} =  [(c_{1}+\dots +c_{n})-P(c_{1}+\dots +c_{n^-})]/n$$ depending on the number $n^-$ of unlucky ones among them. Strategy changes are triggered only if the following conditions hold: for a tax payer $A_j$, with income $c_j$, iff $$[(c_{1}+\dots +c_{n})-P(c_{1}+\dots +c_{n^-})]/n > c_j(1-T)$$
whereas, for a tax evader $A_j$, depending on whether she has been, respectively, lucky or unlucky, iff $$(1-T)(c_{1}+\dots +c_{n})/n > c_j \qquad \text{or} \qquad (1-T)(c_{1}+\dots +c_{n})/n > (1-P)c_j $$}

\AB{The probability ${\cal Q}_c^{p\rightarrow e}\left(\rho(t);p_a,P,T,z\right)$ that a tax-payer, with income $c$, decides to change strategy and starts evading is, then, given by
\begin{equation}
	{\cal Q}_c^{p\rightarrow e}\left(\rho(t);p_a,P,T,z\right) = \sum_{n_e = 1}^{z} \sum_{n_e^- = 0}^{n_e} \xi_e \left(z,n_e,n_e^-;p_a\right) \left[\prod_{k=1}^{n_e-1} \int_{c_-}^{c_+} dc_k~p(c_k)\right] \int_{h_c}^{c_+} dc_{n_e}~p(c_{n_e})
	\label{Q_c_p_e}
\end{equation}
where
\begin{equation}
	\notag h_c\equiv h_c(P,T,n_e,n_e^-) = \frac{c~n_e~(1-T)-(c_1+\dots +c_{n_e-n_e^-})}{1-\Theta\left(n_e^-\right)P}-(c_{n_e-n_e^-+1}+\dots +c_{n_e-1})
	\label{h_c}
\end{equation}
being $\Theta(x)$ the Heaviside step function, which equals $1$ if $x\geq 0$ and $0$ otherwise.} The form taken by $h_c$ follows by considering the $n_e$-th tax-evader as one of the unlucky ones (if any).

\AB{Similarly, the probability ${\cal Q}_c^{e^\pm\rightarrow p}\left(\rho(t);p_a,P,T,z\right)$ that a tax-evader, either lucky ($+$) or unlucky ($-$), changes strategy and starts paying taxes, reads
\begin{equation}
	 {\cal Q}_c^{e^\pm\rightarrow p}\left(\rho(t);p_a,P,T,z\right) = \sum_{n_p = 1}^{z}\xi_p \left(z,n_p\right) \left[\prod_{k=1}^{n_p-1} \int_{c_-}^{c_+} dc_k~p(c_k)\right] \int_{g_c^\pm}^{c_+} dc_{n_p}~p(c_{n_p})
\label{Q_c_e_p}
\end{equation}
where
\begin{equation}
	\notag g_c^\pm\equiv g_c^\pm(P,T,n_p)=\frac{c~n_p~(1-P/2\pm P/2)}{1-T}-(c_1+\dots +c_{n_p-1})
\label{g_c}
\end{equation}
}

\AB{Since the probability of an agent, with income $c$, of being selected for the strategy update in the time interval $dt$ is, respectively, $\rho_c dt$ if she is a tax evader or $(1-\rho_c)dt$ if she is a tax-payer, the approximate mean-value dynamic equation (see Weidlich (1991)) for $\rho_c(t)$, is given by}
\begin{align}
	\notag \frac{d\rho_c(t)}{dt} =&  ~p(c)\left[1-\rho(t)\right] {\cal Q}_c^{p\rightarrow e}\left(\rho(t);p_a,P,T,z\right)\\
	& -p(c)\rho(t)\left[p_a{\cal Q}_c^{e^-\rightarrow p}\left(\rho(t);p_a,P,T,z\right)+\left(1-p_a\right){\cal Q}_c^{e^+\rightarrow p}\left(\rho(t);p_a,P,T,z\right)\right] 
\label{dyn_eq_rho_c}
\end{align}

Integrating over the interval $[c_-,c_+]$, we \AB{finally} get the dynamic equation for $\rho(t)$, \AB{which is}
\begin{equation}
	\frac{d\rho(t)}{dt} =  \int_{c_-}^{c_+} dc \frac{d\rho_c(t)}{dt}
\label{dyn_eq_rho}
\end{equation}

\AB{
Notice that a $\rho$ and a $(1-\rho)$ can be always factorized out from, respectively, ${\cal Q}_c^{p\rightarrow e}$ and ${\cal Q}_c^{e^\pm\rightarrow p}$. Thus the r.h.s. of Eq.~(\ref{dyn_eq_rho}) can be put in the form $\rho(t)\left[1-\rho(t)\right]\left\{\cdots\right\}$, from which it immediately follows $\rho=0,1$ to be absorbing states. The terms in curly brackets, according to Eqs.~(\ref{Q_c_p_e})--(\ref{dyn_eq_rho_c})}, define a polynomial of degree $z-1$ in $\rho$ and of degree $z$ in $p_a$. Therefore, the stationarity condition $d\rho(t)/dt=0$ reduces to solving such polynomial equation, \AB{which yields the stationary solution $\rho^*\equiv \rho^*(p_a)$ as a function of the control parameter $p_a$, needed to derive the expression for critical points and equilibrium stable states.}

\subsection{Critical points and equilibria}
\label{sec:critical}

\AB{We derive a general expression, for any $z$ and any income distribution $p(c)$, for the critical point $p_a^{(e)}$, above which tax-evaders are expected to disappear (i.e., $\rho=0$ is the stable equilibrium). To this end, we impose the stationarity condition $d\rho(t)/dt=0$ in Eq.~(\ref{dyn_eq_rho}) and linearize it by considering $\rho = \epsilon \ll 1$ and neglecting ${\cal O}(\epsilon^2)$ terms. This implies to retain ${\cal O}(1)$ terms in ${\cal Q}_c^{e^\pm\rightarrow p}$ and ${\cal O}(\epsilon)$ terms in ${\cal Q}_c^{p\rightarrow e}$.} With some algebra, one gets a linear expression in $p_a$ providing the following critical point,
\begin{equation}
	p_a^{(e)} \equiv p_a^{(e)}\left(z,p(c)\right) = \frac{\displaystyle\int_{c_-}^{c_+} dc~p(c) \left[z\int_{c(1-T)}^{c_+} dc_1~p(c_1) - \left[\prod_{k=1}^{z-1} \int_{c_-}^{c_+} dc_k~p(c_k)\right]\int_{g_c^+(z)}^{c_+} dc_z~p(c_z)\right]}{\displaystyle\int_{c_-}^{c_+} dc~p(c) \left[z\int_{c(1-T)}^{c\frac{1-T}{1-P}} dc_1~p(c_1) + \left[\prod_{k=1}^{z-1} \int_{c_-}^{c_+} dc_k~p(c_k)\right]\int_{g_c^-(z)}^{g_c^+(z)} dc_z~p(c_z)\right]}
\label{p_a^e}
\end{equation}
where $g_c^\pm(z)\equiv g_c^\pm(P,T,z)$.

Linearizing instead Eq.~(\ref{dyn_eq_rho}) around $\rho = 1$ \AB{(i.e., considering $\rho = 1-\epsilon$, with $\epsilon \ll 1$), and retaining only ${\cal O}(\epsilon)$ terms, the following polynomial equation of degree $z$ in $p_a$ is obtained}
\begin{align}
	\notag 0 =& \sum_{n_e^-=0}^{z}\binom{z}{n_e^-}\left(1-p_a\right)^{z-n_e^-}p_a^{n_e^-}\int_{c_-}^{c_+} dc~p(c) \left[\prod_{k=1}^{z-1} \int_{c_-}^{c_+} dc_k~p(c_k)\right]\int_{h_c(z)}^{c_+} dc_z~p(c_z) ~+ \\
	&- z\int_{c_-}^{c_+} dc p(c)\left[p_a\int_{c\frac{1-P}{1-T}}^{c\frac{1}{1-T}} dc_1~p(c_1) + \int_{c\frac{1}{1-T}}^{c_+} dc_1~p(c_1)\right] 
\label{p_a^p}
\end{align}
where $h_c(z)\equiv h_c(P,T,z,z)$. Solving Eq.~(\ref{p_a^p}) one finds the critical point $p_a^{(p)}$ below which tax-payers are estimated to disappear (i.e., $\rho=1$ is the stable equilibrium).

A stability analysis reveals that for $p_a^{(p)} < p_a < p_a^{(e)}$ the states $\rho=0$ and $\rho=1$ become unstable, while the stable equilibrium $\rho^*\in\left(0,1\right)$ is a polymorphic state, which is solution of a polynomial equation of degree $z-1$ in $\rho$. For $z=1$, the latter is of degree zero in $\rho$ and linear in $p_a$. \AB{By solving it, the first-order transition from $\rho^*=1$ to $\rho^*=0$ is found at a value of $p_a=p_a^{(e)}=p_a^{(p)}$ depending on $T$, $P$ and $p(c)$. This can be obtained by taking $z=1$ in Eq.~(\ref{p_a^e}) or (\ref{p_a^p}). For $z=2$, instead, the stationarity condition reduces to a linear equation in $\rho$ and the equilibrium $\rho^*$ can be expressed in terms of $p_a$, as}
\begin{equation}
	\rho^*(p_a) = \frac{\sum_{a=1}^{4} {\cal I}_a}{\sum_{a=1}^{9} {\cal I}_a}
\label{rho*_2}
\end{equation}
where $\{{\cal I}_a\}_{a\in\{1,\dots,9\}}$ are nine integrals, listed in Appendix. It is worth to note that $\rho^*(p_a)$ describes a second-order phase transition connecting continuously $\rho^*=1$ at $p_a^{(p)}$ to $\rho^*=0$ at $p_a^{(e)}$. \AB{If $p_a^{(p)}$ and/or $p_a^{(e)}\notin\left[0,1\right]$, we estimate $\rho=0$ and/or $\rho=1$ are always unstable equilibria.}

For $z\geq 3$ the computation of $\rho^*(p_a)$ becomes too long to be amenable to a by-hand calculation. Besides, we expect $\rho^*(p_a)$ to describe a continuous phase transition and, therefore, no qualitative changes with respect to the case $z=2$. \AB{Contrarily, changing from $z=1$, i.e., interaction with just another agent at time, to $z=2$, i.e.,  interaction with with two agents at time, modifies the phase transition, passing from discontinuous to continuous.}

\section{\AB{Results for a well-mixed population}}
\label{sec:results}

\AP{In this section we present and discuss the results provided by the theoretical analysis of the PTM for a well-mixed population, testing the analytical predictions through numerical Monte Carlo (MC) simulations. We first consider the simplest, limit case of individuals getting all the same income (\lq\lq egalitarian" population). This perfectly homogeneous case plays here the role of a null model to which any heterogeneous case should be compared.} 

\AB{The perfect homogeneity is implemented by assuming that every agent has the same income $\bar c$ and considering $p(c)=\delta(c-\bar c)$, where $\delta(x)$ is the delta distribution. Further, it will shown below that the value of $\bar c$ plays no role. In this case the analytical tractability is significantly simplified. Indeed, lucky tax-evaders never change strategies, while unlucky ones certainly do whenever at least one tax-payer is present among the $z$ chosen agents. Therefore, ${\cal Q}^{e^+\rightarrow p}=0$ and ${\cal Q}^{e^-\rightarrow p}=1-\rho^z$. Also ${\cal Q}^{p\rightarrow e}$ has now a simpler expression, which is}
\begin{equation}
	{\cal Q}^{p\rightarrow e}\left(\rho(t);p_a,P,T,z\right) = \sum_{n_e = 1}^{z} \sum_{n_e^- = 0}^{\left\lfloor n_eT/P\right\rfloor}\binom{z}{n_e}\binom{n_e}{n_e^-}\left(1-p_a\right)^{n_e-n_e^-}p_a^{n_e^-}\left[\rho(t)\right]^{n_e}\left[1-\rho(t)\right]^{z-n_e}
\label{Q_p_e_delta}
\end{equation}
where $\left\lfloor x\right\rfloor$ is a floor function assigning to $ x\in\mathbb{R}$ the greatest integer less than $x$. \AB{For a given value of $z$ and once fixed $p_a$, ${\cal Q}^{p\rightarrow e}$ --and hence the dynamics-- depends on $T$ and $P$ through only their ratio $T/P$ and in a discrete way. Both these characteristics will not hold in the heterogeneous income case, as shown below in the Appendix.}

The tax-evaders extinction point is simply given by $p_a^{(e)}=z/(z+1)$, independently from $P$ and $T$. For $z=1$, in particular, we find the first-order transition located at $p_a^{(e)}=p_a^{(p)}=1/2$. For $z>1$, $p_a^{(p)}$ ($<p_a^{(e)}$) is found as solution of Eq.~(\ref{p_a^p}), whose form becomes
\begin{equation}
	0 = z p_a - \sum_{n_e^-=0}^{\left\lfloor zT/P\right\rfloor}\binom{z}{n_e^-}\left(1-p_a\right)^{z-n_e^-}p_a^{n_e^-}
\label{p_a^p_delta}
\end{equation}
Given $z$, depending on the value of $T/P<1$ we get different values for $p_a^{(p)}$. In particular, for $z=2$, Eq.~(\ref{p_a^p_delta}) gives $p_a^{(p)}=\sqrt{2}-1$ for $P/2 < T < P$ and $p_a^{(p)}=2-\sqrt{3}$ for $T \leq P/2$, both lower than $p_a^{(e)}=2/3$.

According to Eq.~(\ref{rho*_2}), for $z=2$, the stable equilibrium is given by $\rho^*(p_a) = (2-3p_a)/(1-p_a+p_a^2)$ for $P/2 < T < P$ and $\rho^*(p_a) = (2-3p_a)/(1+p_a-p_a^2)$ for $T \leq P/2$, provided $p_a^{(p)} \leq p_a \leq p_a^{(e)}$. Consistently, $\rho^*\left(p_a^{(e)}\right) = 0$ and $\rho^*\left(p_a^{(p)}\right) = 1$. 

\AB{In real socio-economic systems, wealth and income are far from being uniformly distributed. Self-reinforcing dynamics -- of the type `rich-gets-richer`-- make the wealth and income distributions heavy-tailed, thus leading to an overwhelming majority of poor individuals and to a small minority of very rich ones. Then, it is also worth to consider the case of power-law distributed incomes (\lq\lq aristocratic" population).}

\begin{figure}[tb!]
	\centering
		\includegraphics[width=.49\linewidth]{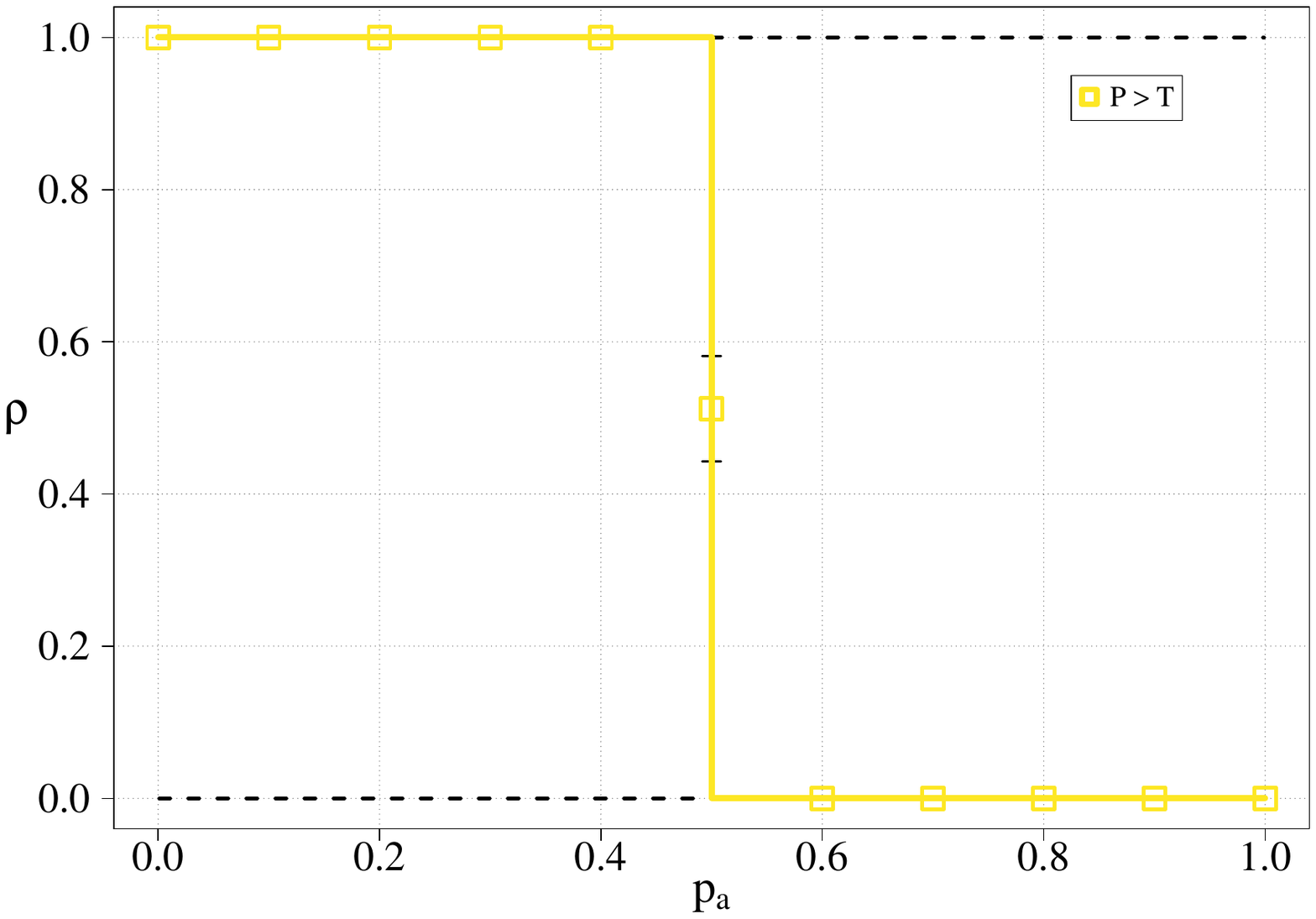}
		\hfill
		\includegraphics[width=.49\linewidth]{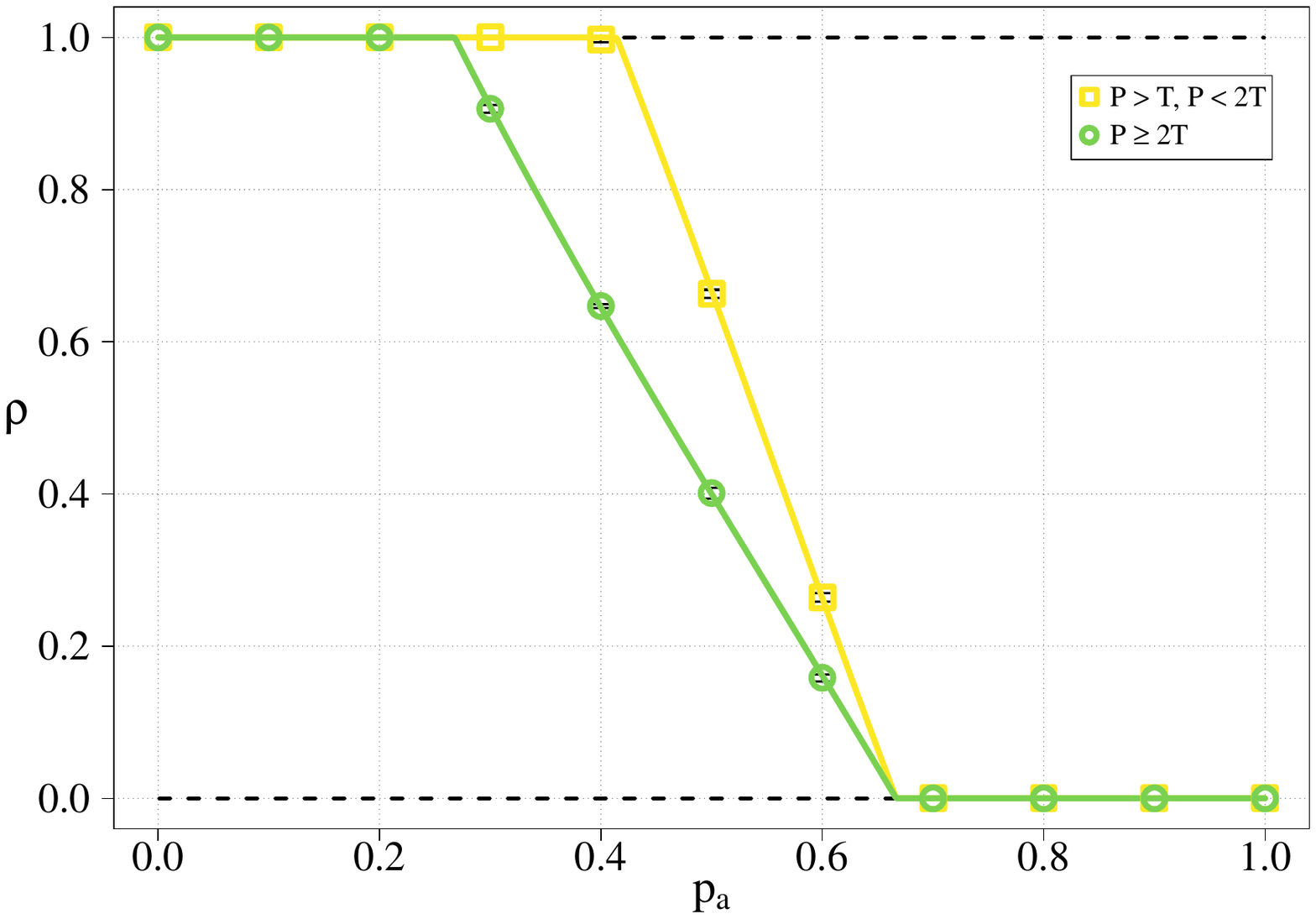}
		\\
		\includegraphics[width=.49\linewidth]{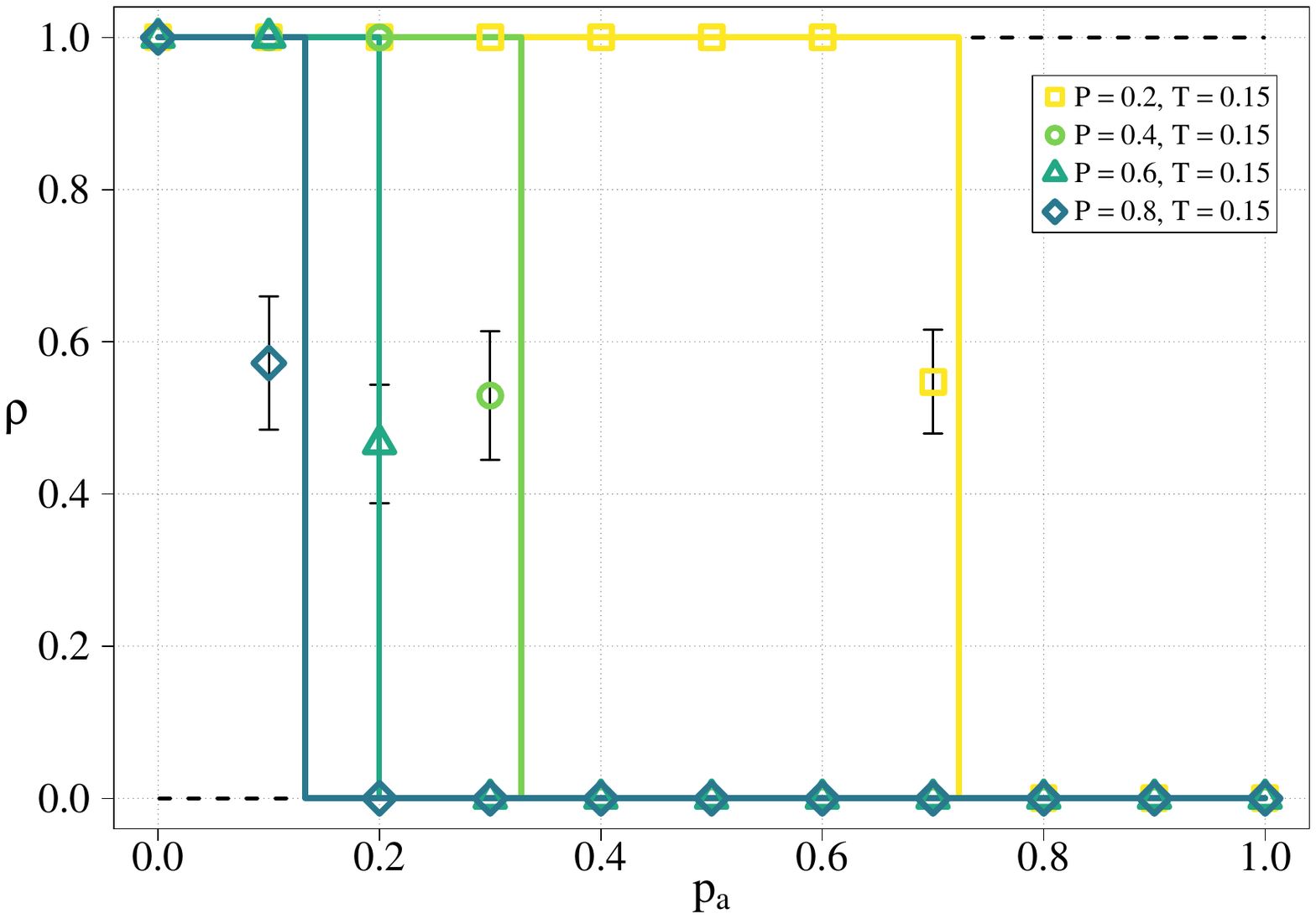}
		\hfill
		\includegraphics[width=.49\linewidth]{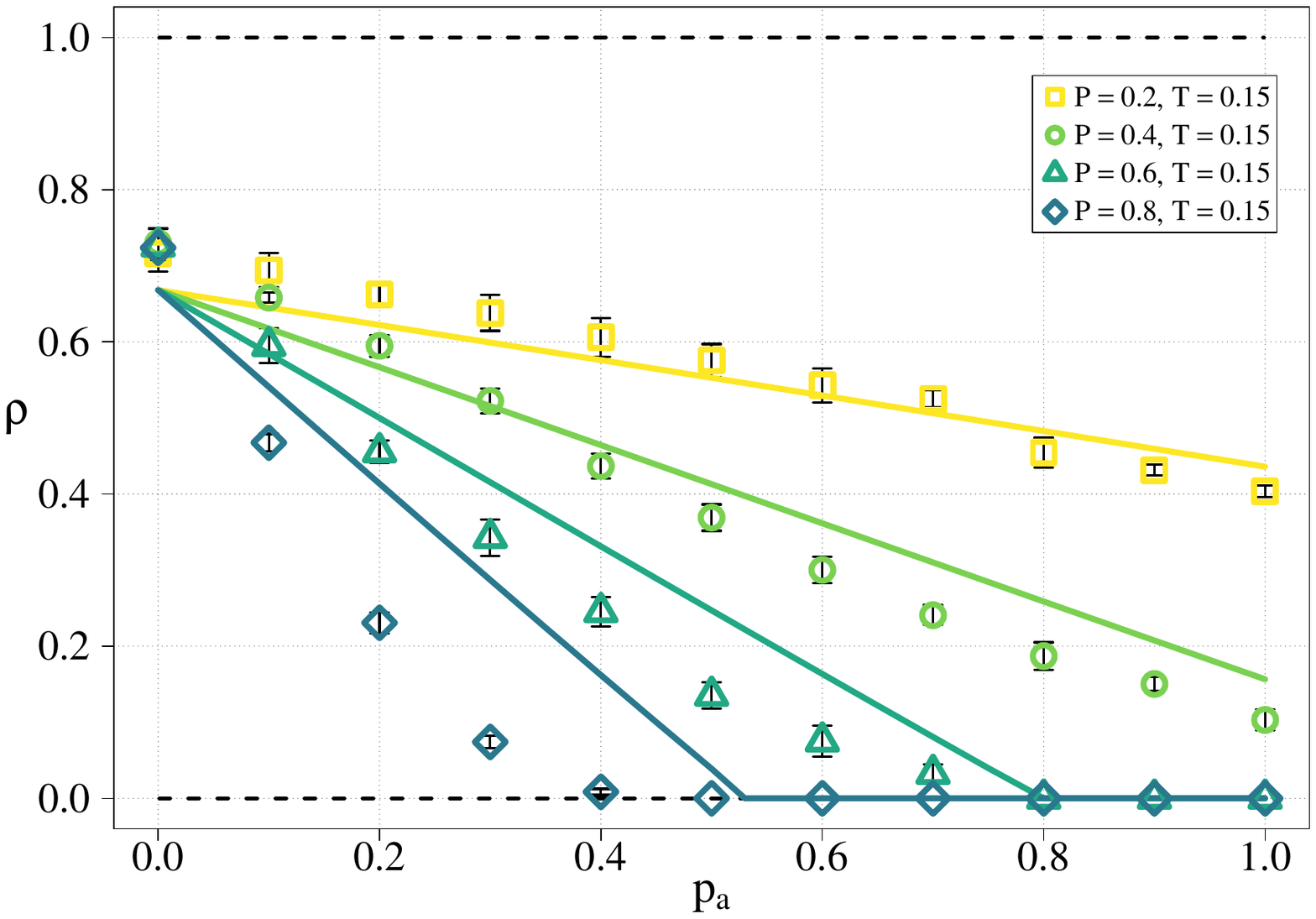}
		
		\caption{\AP{Results for a well mixed population. Stable equilibrium proportion of tax-evaders $\rho$ versus $p_a$, for $z=1$ (left panels) and $z=2$ (right panels) for different income distributions. Points represent the MC's results and bars the respective standard deviations, while solid lines represent the theoretical predictions. Dashed lines mark the predicted unstable equilibria. {\em Top panels.} Results for a delta income distribution, that holds for any $P>T$. {\em Bottom panels.} Results for a power-law distribution. We set here $T=0.15$, while different values of the fixed proportional fine $P$ are considered. Parameters for the distribution are $\kappa=1.6$, $c_-=1$ and $c_+=100$. }
		}
\label{fig:results}
\end{figure}

\AB{The normalized power-law distribution $p_\kappa(c)$ with exponent $\kappa>1$ and mode $c_-$, is given by}
\begin{equation}
	p_\kappa(c) = \Theta\left[\left(c-c_-\right)\left(c_+-c\right)\right]\frac{\left(1-\kappa\right)c^{-\kappa}}{{c_+}^{1-\kappa}-{c_-}^{1-\kappa}}
\label{power-law_dist}
\end{equation}

\AP{The equilibrium solutions for $z=1$ and $z=2$ as function of the audit probability, computed from Eqs.~(\ref{p_a^e}) and (\ref{rho*_2}), are shown in Figure~\ref{fig:results}   for both the delta and the power-law distributions of income. Points and bars represent simulation results, with the respective standard deviations, while solid lines represent theoretical predictions. As expected, in the case of the delta distribution (top panels) results do not depend on specific values of $T$ and $P$ (provided that $P>T$), thus only one curve appears for any $z$. On the contrary, in the case of the power-law distribution (bottom panels), results are presented for $T=0.15$ and the different curves correspond to different values of the proportional fine $P$ (the explicit form of the first-order transition point $p_a^{(e)}=p_a^{(p)}$ for $z=1$ is given in Appendix).}

Looking at the figure, it it evident that the theoretical predictions are very accurate when all the individuals get the same income, whereas they loose accuracy when individuals get different incomes. The reason for this is to be found in the strong assumption we made at the beginning: considering strategy and income as uncorrelated variables at any time. \AB{We know this can be true only in first rounds, when strategies and incomes are still mainly assigned at random. As time passes by and the equilibrium is reached, some correlations between strategy and income develops, making the outcome deviate from the theoretical prediction.}

\begin{figure}[tb!]
	\centering
		\includegraphics[width=.49\linewidth]{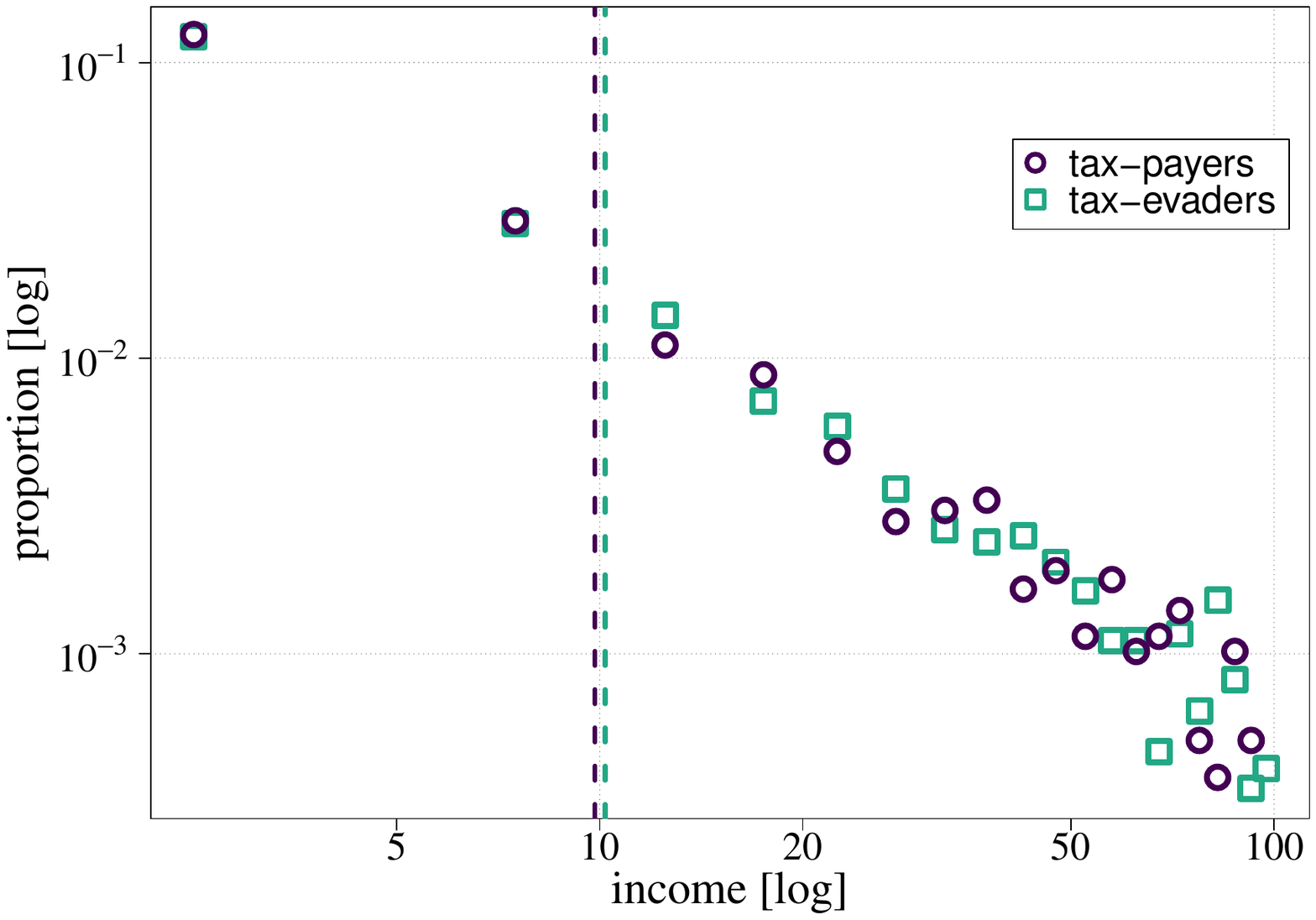}
		\hfill
		\includegraphics[width=.49\linewidth]{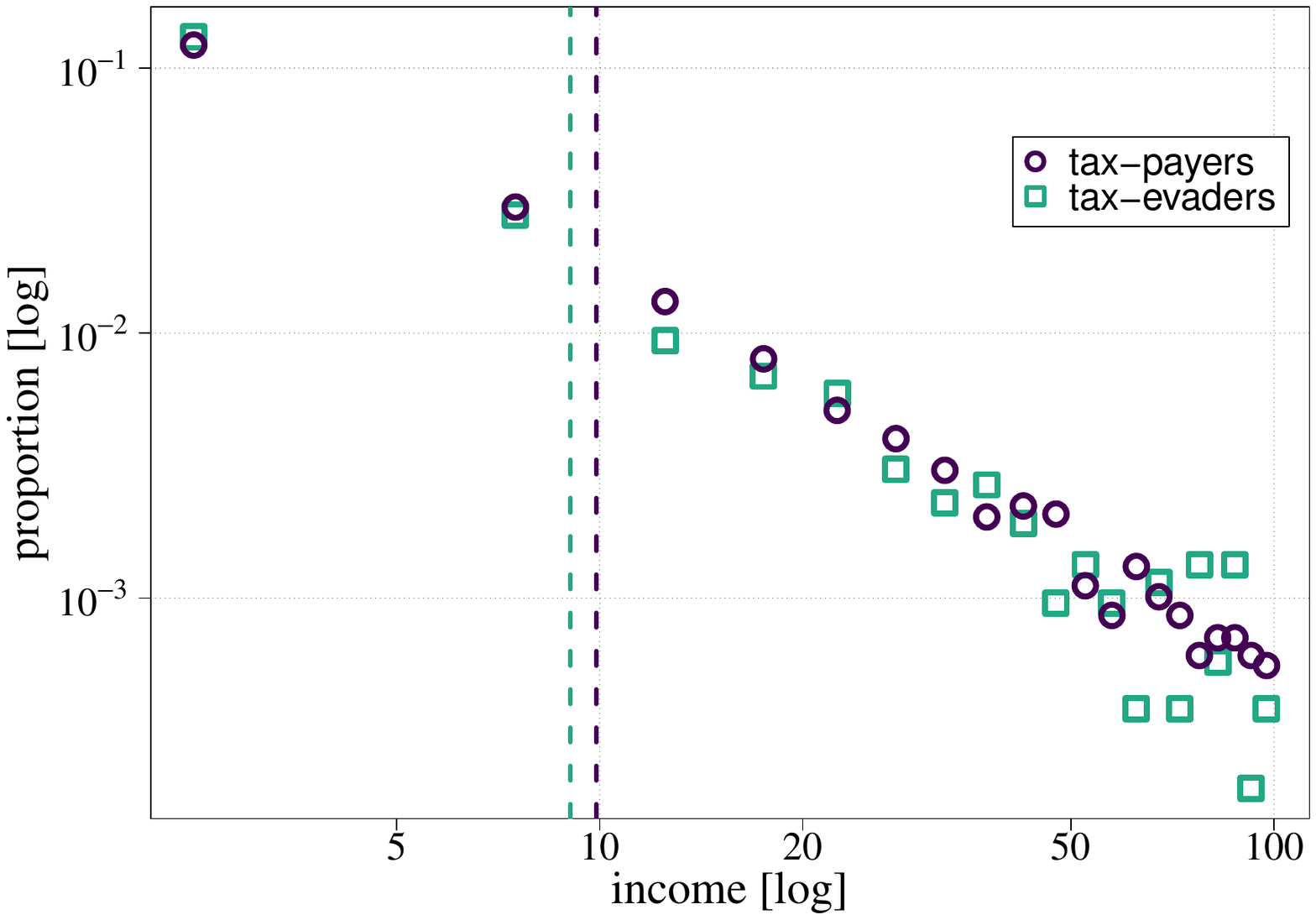}
		
		\caption{Tail distributions of the income at the equilibrium for the two groups of strategists, for $p_a=0.1$ (left) and $p_a=0.7$ (right), as found in two different runs for the same power-law distribution of Figure~\ref{fig:results}, taking $P=0.4$ and $T=0.15$. Vertical dashed lines indicate the respective average incomes.
		}
\label{fig:histograms}
\end{figure}

Focusing, for instance, on the case $z=2$ (right bottom panel of Figure~\ref{fig:results}), we report, in Figure~\ref{fig:histograms}, the tail distributions of the income (i.e., the probability to get an income greater or equal to a certain value) at the equilibrium for the two groups of strategists, showing two realizations, one for $p_a=0.1$ and one for $p_a=0.7$. If there were no correlation between strategy and income, we would find the same distribution for both groups. Instead, we find the tax-evaders (tax-payers) income distribution to be more pronounced towards high incomes than the tax-payers (tax-evaders) in the first (second) case in which $\rho_{MC}\approx 0.66$ versus the predicted $\rho\approx 0.62$ ($\rho_{MC}\approx 0.24$ versus the predicted $\rho\approx 0.31$). To better understand the relevance of such correlations, it has to be noted that, because of the well-mixing, giving the chance to any agent to interact with any other agent, even small differences between the two income distributions can have a significant impact on the equilibrium state.

That said, coming again to Figure~\ref{fig:results}, the proportional fine turns out to be an effective policy, able to strongly contrast the spreading of the cheating strategy, for both income distributions. Remarkably, \AB{a strong enough fine $P$ comes out to fully unravel tax evasion.} \AP{By the way, we find that the minimal value of $p_a$ sufficient to reach this goal is lower for the power-law income distribution, provided that $P$ is high enough. For $T=0.15$, this holds above $P=0.2775$ for $z=1$, and above $P\approx 0.7$ for $z=2$. Of practical interest is also observing that, for the power-law income distribution, the proportion of tax-evaders stays definitely below $1$ even for low values of $p_a$. This resilience against tax evasion is due to the tail of high-income tax-payers. The same holds for the high-income tax-evaders when $p_a$ takes high values, but only if $P$ is not too larger than $T$; otherwise, soon or later, each individual will prefer to pay taxes.}

\subsection{\AB{Time-dependent audit probability}}
\label{sec:p_a(t)}

An interesting observation regards the cost entailed by audits. \AB{Costs of audits are proportional to their frequency and accuracy, and are paid by means of tax revenues obtained by all citizens}, either being they public servants, private employees or employers, or freelancers. While public servants can only play a marginal role in the studied problem, private employers and freelancers are those who can take great advantage from evading taxes. In other words, it is mainly the presence of the latter category to make strictly necessary an efficient (and hence costly) supervision. Therefore, given a real, entire population, we can consider the modeled population as that subset consisting of those citizens for which the dilemma of tax evasion is relevant. Starting from this observation, we propose a refinement of the model intended to cover a fraction of the supervision cost by only means of the contributions (through taxes and/or fines) coming from the modeled population (i.e., private employers and freelancers).

A simple way of implementing this idea is by considering a time-dependent audit probability $p_a(t)$ and coupling it to the global state $\rho(t)$ of the population. We let $p_a$ evolve as a function of the average contribution of an agent, $\bar C(t)$. \AB{Since an agent, with income $c$, will either pay the tax amount $cT$ or the fine $cP$, the expected value for the individual contribution, $\bar C(t)$, is}
\begin{equation}
	\bar C(t) = \int_{c_-}^{c_+} dc~p(c) \left[T\left[1-\rho(t)\right]+P p_{a}(t)\rho(t)\right] c = \left[T\left[1-\rho(t)\right]+P p_{a}(t)\rho(t)\right] \bar c
	\label{C(t)}
\end{equation} 
Specifically, we assume $p_a(t)$ to obey the following recursive relation
\begin{equation}
	p_{a}(t) = \tilde p_{a} + (1-\tilde p_{a})\beta~{\cal F}\left(\bar C(t)\right)
	\label{p_a(t)}
\end{equation}
where ${\cal F}$ is an increasing function bounded in $[0,1]$ and such that ${\cal F}(x=0)=0$; \AB{$\tilde p_{a}$ is the level of supervision maintained by the government regardless of agents' contribution;} $\beta\in[0,1]$ weighs $\bar C(t)$ and hence, indirectly, the time-dependence of $p_{a}(t)$. \AB{The model with time-independent $p_a$ is obtained as a particular case of $\beta=0$. Eq.~(\ref{dyn_eq_rho}) is here understood taking $p_a \equiv p_a(t)$.} 

\AB{A quite general choice for ${\cal F}$ is a power function, ${\cal F}(x)=x^{\gamma}/{\bar c}$, with $ \gamma\in \mathbb{Q}$ to make analytically tractable the stability analysis.} The normalization with respect to $\bar c$ makes ${\cal F}(x)\in[0,1]$, on one hand, and frees the coupling between $p_a$ and $\rho$ from any dependence on the specific value of $\bar c$, on the other. Depending on the value of $ \gamma$, the supervision cost depends sub-linearly ($\gamma<1$), linearly ($\gamma=1$), or super-linearly ($\gamma>1$) on the contribution $\bar C$. Being ${\bar C}/{\bar c}\in\left[0,1\right]$, the sub-linear (super-linear) case weighs more (less) $\bar C$ than the linear case, hence favoring more (less) the \textsc{pay} strategy.

\AB{Critical values of $\tilde{p}_a$, $T$, and $P$, are derived in the next subsection. Moreover, same considerations about the birth of some correlations between strategy and income, already done in Section~\ref{sec:results}, hold identical here.}

\begin{figure}[tb!]
	\centering
		\includegraphics[width=.49\linewidth]{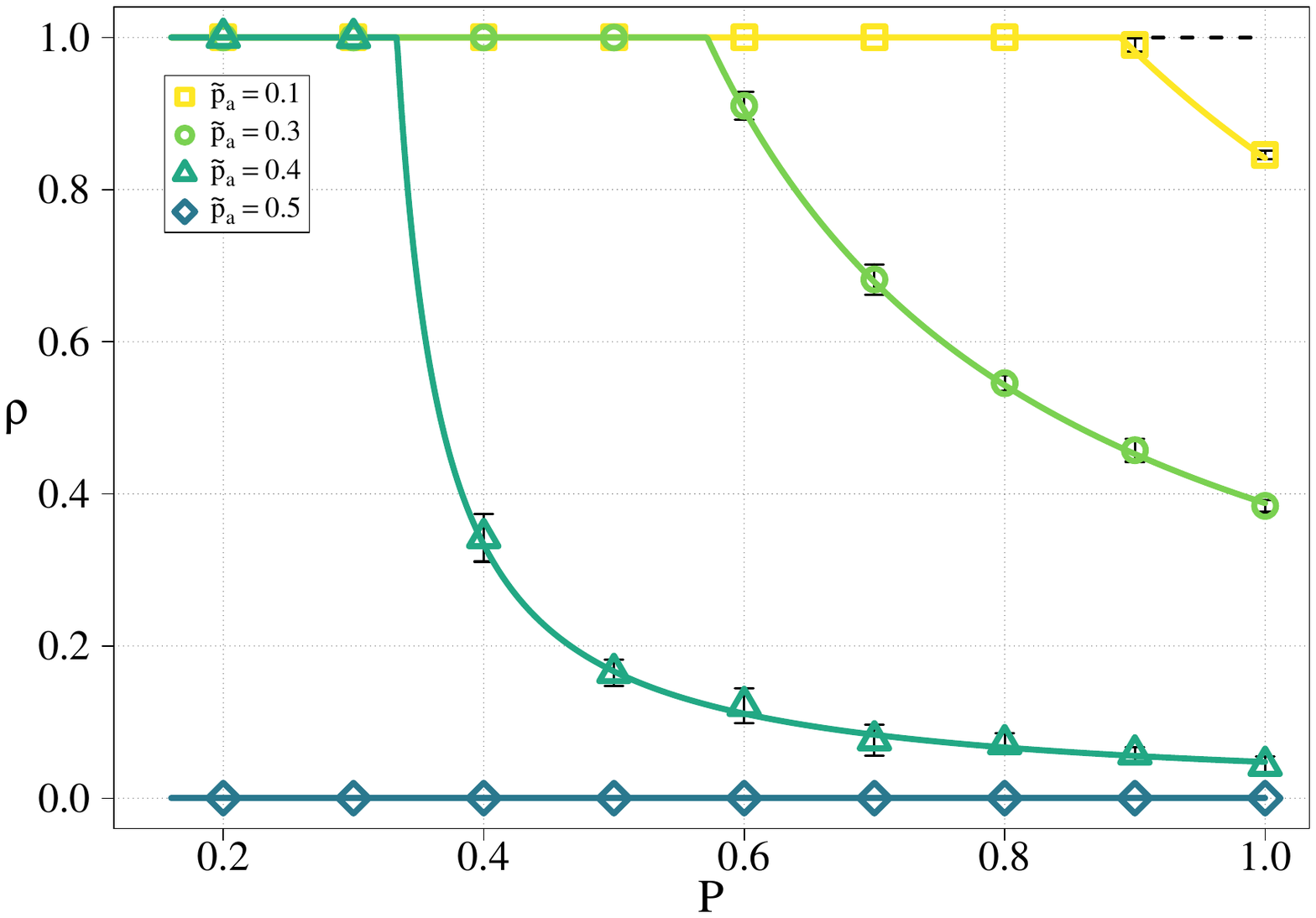}
		\hfill
		\includegraphics[width=.49\linewidth]{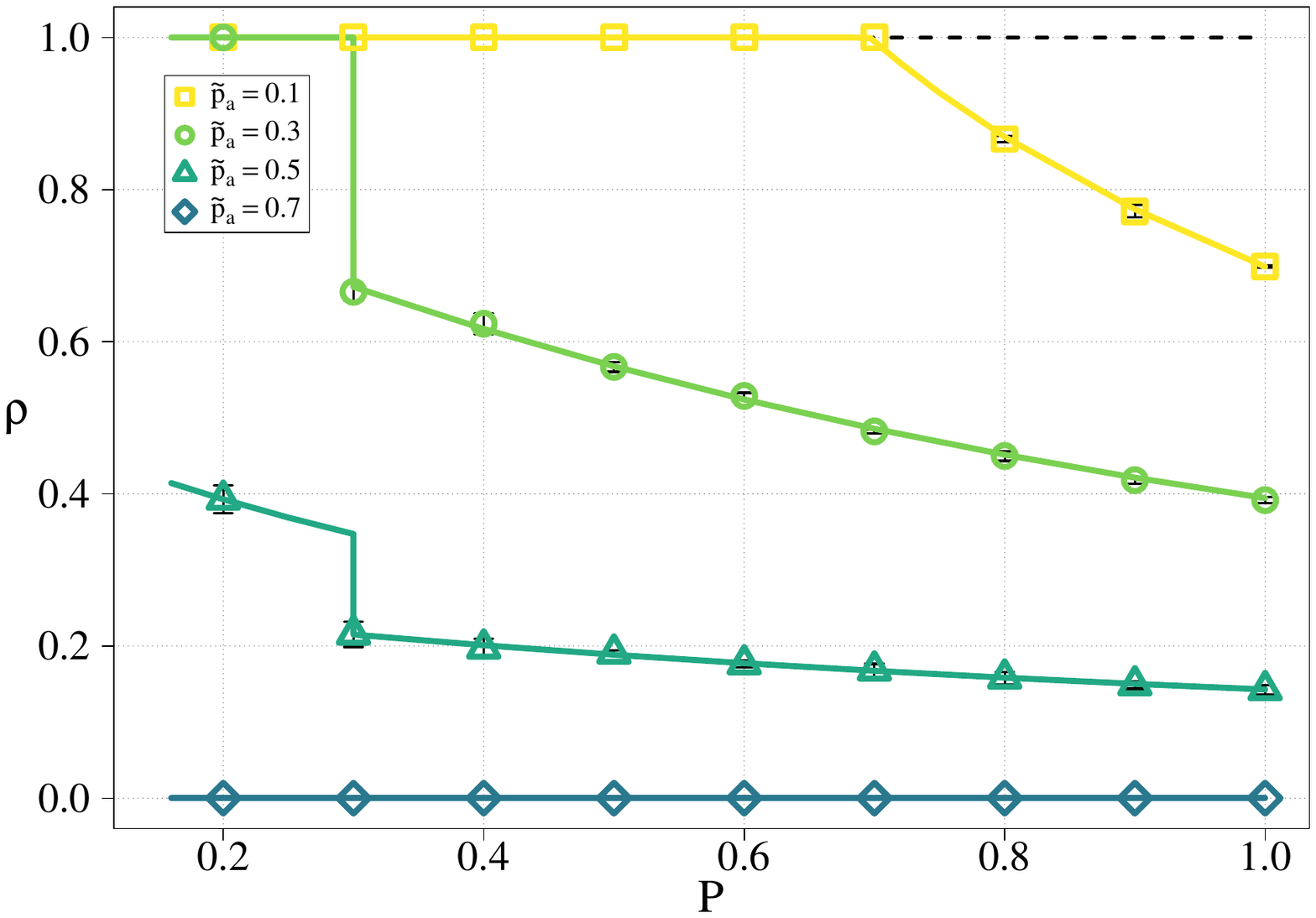}
		\\
		\includegraphics[width=.49\linewidth]{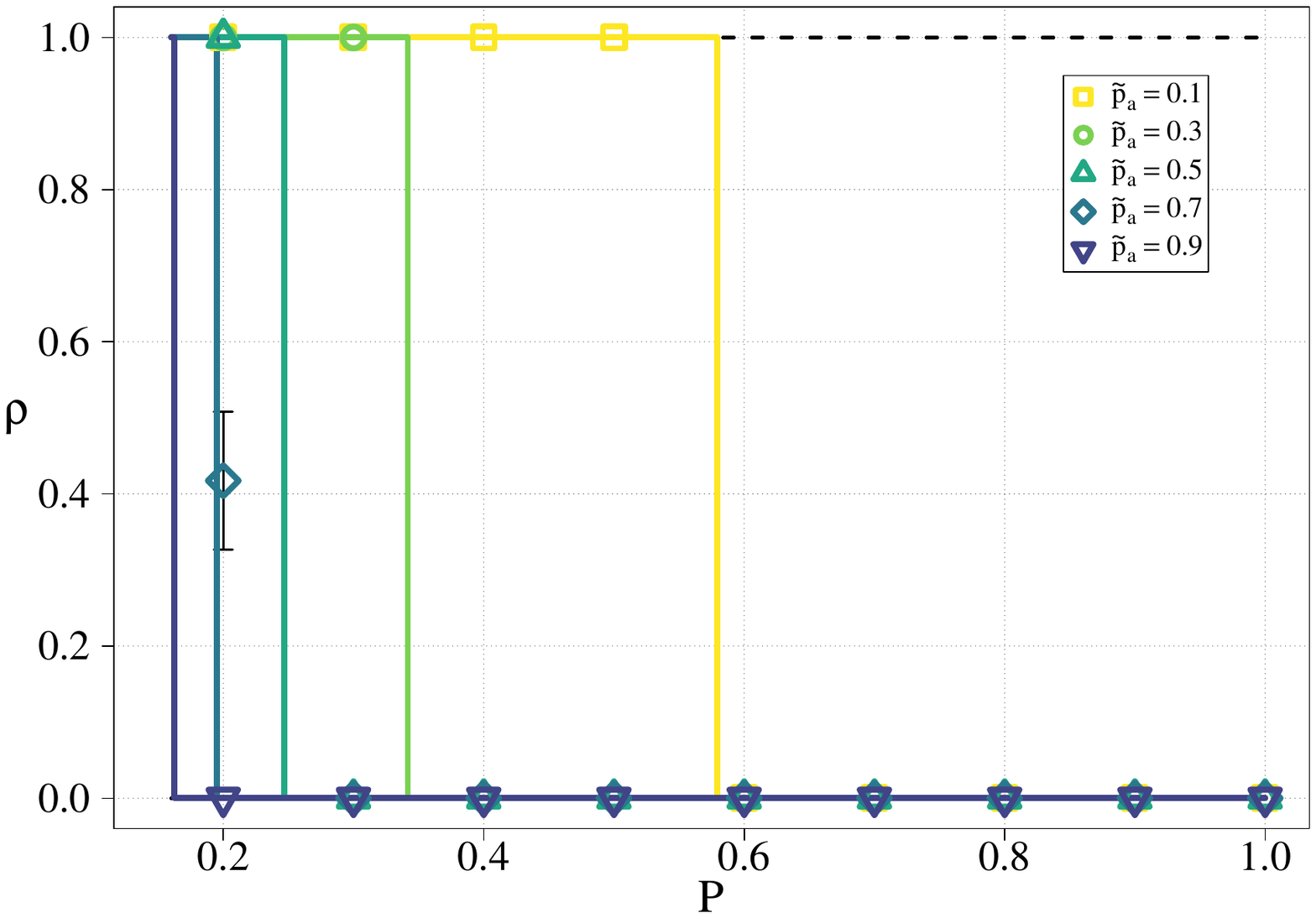}
		\hfill
		\includegraphics[width=.49\linewidth]{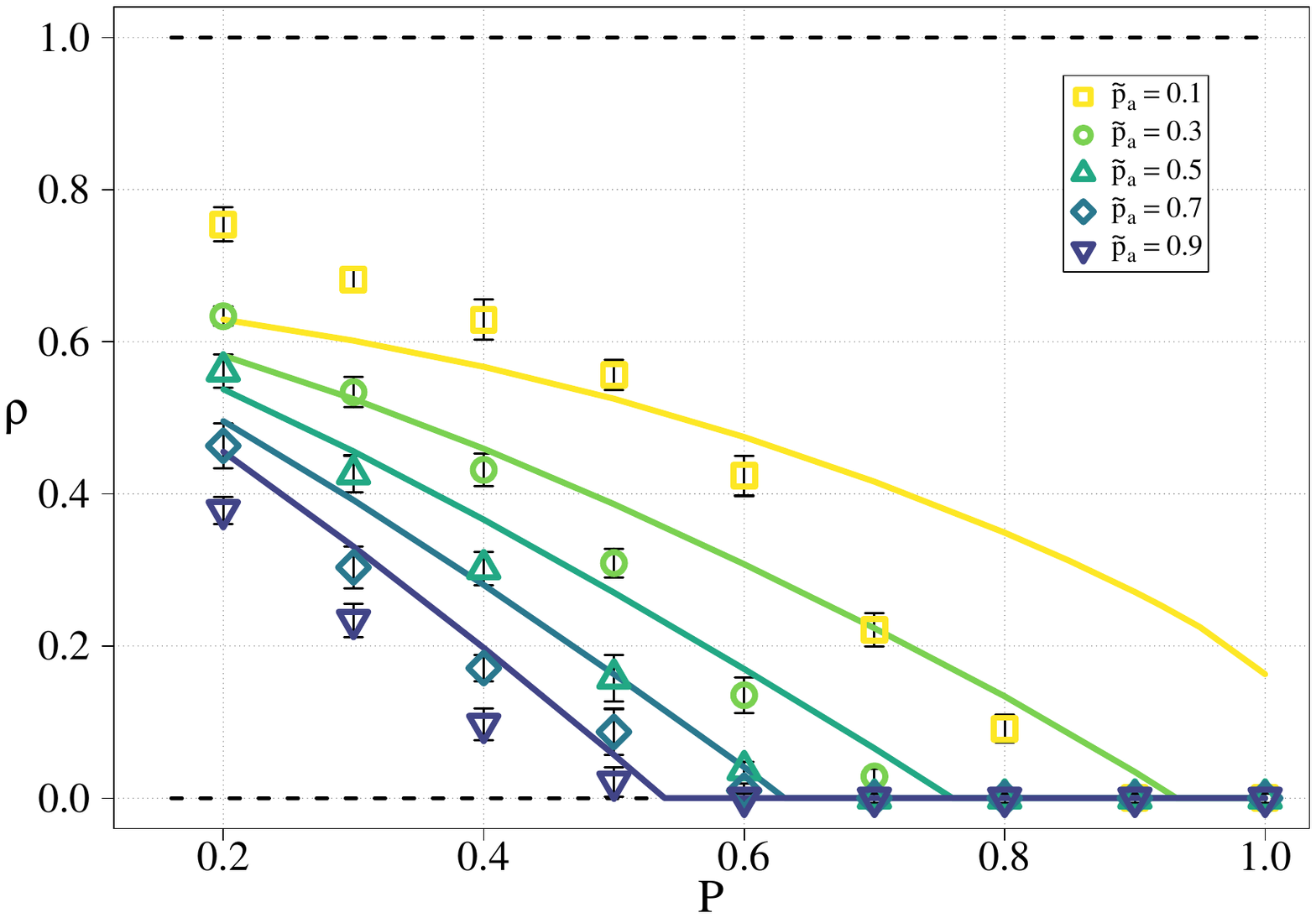}
		
		\caption{\AB{Results for time-dependent audit probability. Stable equilibrium proportion of tax-evaders $\rho$ versus $P$, for $z=1$ (left) and $z=2$ (right); $T=0.15$, $\beta =1$, $\gamma = 1$. Points depict Monte Carlo results, bars stand for respective standard deviations, while solid lines represent theoretical predictions. Dashed lines mark the predicted unstable equilibria. {\em Top panels.} Results for the delta distribution. The jump observed for $z=2$ at $P=0.3$ is due to the discrete change in ${\cal Q}^{p\rightarrow e}$ at $P=2T$ (see Eqs.~(\ref{Q_p_e_delta})--(\ref{p_a^p_delta})). {\em Bottom panels.} Results for a power-law distribution with $\kappa=1.6$, $c_-=1$ and $c_+=100$.}}
\label{fig:results_p_a(t)}
\end{figure}

In Figure~\ref{fig:results_p_a(t)} we show the equilibrium values of $\rho$ for $T=0.15$, $\beta =1$, and $\gamma = 1$. To measure the effect of the refinement we made with Eq.~(\ref{p_a(t)}), we compare the value of $\rho$ at the equilibrium, obtained taking the same value for the time-independent $p_a$ and the baseline $\tilde{p}_a$, for different values of $P$ (i.e., the points on the different curves for a fixed $p_a$, on one hand, and the points on the curve corresponding to $\tilde{p}_a=p_a$, on the other). \AB{A direct comparison between Figure~\ref{fig:results} and Figure~\ref{fig:results_p_a(t)}, shows that the control policy always performs better after the refinement, except in case of very low values of $P$ and $\tilde{p}_a$ with the power-law income distribution, in which results are approximately invariant. Interestingly, however, in the latter case, for both $z=1$ and $z=2$, high enough values of $P$ ensure the extinction of tax-evaders, for all explored values of $\tilde{p}_a$. Instead, for the delta income distribution, evaders survive until $\tilde{p}_a$ is sufficiently high.}

\subsubsection{Critical Points and Equilbria}
\label{sec:equilibria_p_a(t)}

In this subsection, we firstly derive the values of $\tilde p_{a}$ above and below which, respectively, tax-evaders and tax-payers are expected to be extinct, for given values of $T$ and $P$.

When linearizing Eq.~(\ref{dyn_eq_rho}) around $\rho = 0$ or $\rho = 1$, $p_a(t)$ contributes with only its ${\cal O}(1)$ terms. \AB{Thus, from Eqs.~(\ref{C(t)})--(\ref{p_a(t)}), in the case in which $\rho^\ast = 0$, $p_a$ converges to
\begin{equation}
	p_a^\ast =\tilde p_{a} + (1-\tilde p_{a})\beta T^\gamma
	\label{p_a*_rho0}
\end{equation} 
Instead, when $\rho^\ast = 1$, $p_a^\ast $ is obtained as the solution of
\begin{equation}
	\left[(1-\tilde p_{a})\beta P^\gamma\right]{p_a}^\gamma - p_a + \tilde p_{a}= 0
	\label{p_a*_rho1}
\end{equation} 
}
Now, the solutions for $p_a$ found in Section~\ref{sec:critical} are still valid and must coincide with those yield by Eqs.~(\ref{p_a*_rho0}) and (\ref{p_a*_rho1}). Precisely, the value of $p_a^\ast$ given by Eq.~(\ref{p_a*_rho0}) (Eq.~(\ref{p_a*_rho1})) matches the critical value $p_a^{(e)}$ ($p_a^{(p)}$) provided by Eq.~(\ref{p_a^e}) (Eq.~(\ref{p_a^p})). Therefore, expressing $\tilde p_{a}$ in terms of $p_a^\ast$ through Eq.~(\ref{p_a*_rho0}) (Eq.~(\ref{p_a*_rho1})) and substituting for the latter the value of $p_a^{(e)}$ ($p_a^{(p)}$) found for time-independent $p_a$, we get the critical value of $\tilde p_{a}$ above (below) which tax-evaders (tax-payers) are predicted to be extinct, as a function of $T$ ($P$) and regardless of $P$ ($T$):
\begin{align}
	&\tilde {p}_a^{(e)} = \frac{p_a^{(e)}-\beta T^\gamma}{1-\beta T^\gamma}
	\label{tilde_p_a^e} \\
	&\tilde {p}_a^{(p)} = \frac{p_a^{(p)}-\beta \left(p_a^\ast P\right)^\gamma}{1-\beta \left(p_a^\ast P\right)^\gamma}
	\label{tilde_p_a^p}
\end{align}

Alternatively, from the same expressions, one can find the critical values of $T$ and $P$ in terms of $\tilde p_{a}$, as follows
\begin{align}
	&T^{(e)} = \left[\frac{p_a^{(e)}-\tilde p_{a}}{\left(1-\tilde p_{a}\right)\beta}\right]^{\displaystyle\frac1{\gamma}}
	\label{T_e} \\
	&P^{(p)} = \frac1{p_a^{(p)}}\left[\frac{p_a^{(p)}-\tilde p_{a}}{\left(1-\tilde p_{a}\right)\beta}\right]^{\displaystyle\frac1{\gamma}}
	\label{P_p}
\end{align} 
Tax-evaders are extinct for $T>T^{(e)}$, $\forall P$, while tax-payers are extinct for $P<P^{(p)}$, $\forall T$. To note that the values of $p_a^{(e)}$ and $p_a^{(p)}$ computed for time-independent $p_a$ generally depends on the values chosen for $T$ and $P$. Therefore, one first computes $p_a^{(e)}$ and $p_a^{(p)}$ for given $T$ and $P$, and then, provided $T^{(e)}$ and $P^{(p)}$ through Eqs.~(\ref{T_e})--(\ref{P_p}), compares the chosen $T$ and $P$ with the respective found critical values. In the special case of equal incomes, however, $p_a^{(e)}$ is independent of $P$ and $T$, while $p_a^{(p)}$ only depends on the ratio $T/P$. Referring specifically to Figure~\ref{fig:results_p_a(t)}, this simplification allow us to derive the value of the tax-evaders extinction point, that, for $T=0.15$, $\beta =1$, and $\gamma = 1$, is given by $\tilde {p}_a^{(e)} \approx 0.412$ for $z=1$ ($p_a^{(e)} = 1/2$), and by $\tilde {p}_a^{(e)} \approx 0.608$ for $z=2$ ($p_a^{(e)} = 2/3$). \AB{Consistently, $\rho$ vanishes for any $\tilde {p}_a >\tilde {p}_a^{(e)}$.}

Remarkably, from Eqs.~(\ref{tilde_p_a^e})--(\ref{tilde_p_a^p}), it follows that $\tilde p_{a}^{(e)}$ ($\tilde p_{a}^{(p)}$) is strictly smaller than $p_a^{(e)}$ ($p_a^{(p)}$) computed for time-independent $p_a$, if and only if $p_a^{(e)}<1$ ($p_a^{(p)}<1$), no matter the values of the other parameters. In other words, whenever $T$, $P$, $\beta$ and $\gamma$ are such that, for time-independent $p_a$, the critical points $p_a^{(e)}$ and $p_a^{(p)}$ take feasible values (i.e., within $[0,1]$), then we are guaranteed to find smaller values for $\tilde p_{a}^{(e)}$ and $\tilde p_{a}^{(p)}$, respectively, hence a more effective control policy. We have seen an instance of this in the previously analyzed example.

Looking now for equilibrium polymorphic populations, we focus on the cases $z=1$ and $z=2$. To this end, we invert Eq.~(\ref{p_a(t)}) to get an expression for $\rho(t)$ in terms of $p_a(t)$,
\begin{equation}
	\rho(t) = \frac{\displaystyle\left[\frac{p_a(t)-\tilde p_{a}}{\left(1-\tilde p_{a}\right)\beta}\right]^{\displaystyle\frac1{\gamma}}-T}{P p_{a}(t)-T}
	\label{rho*_p_a(t)}
\end{equation}

For $z=1$, from the stationarity condition $d\rho(t)/dt=0$, we get a linear equation in $p_a(t)$, giving $p_a^\ast$ equal to the r.h.s of Eq.~(\ref{p_a^e_z=1}) (see Appendix). Substituting in Eq.~(\ref{rho*_p_a(t)}), we get the equilibrium solution $\rho^\ast$.

For $z=2$, the stationarity condition yields the equilibrium solution $\rho^\ast$ given in Eq.~(\ref{rho*_2}), with $p_a(t) = p_a^\ast$. Equating it with the $\rho^\ast$ coming from Eq.~(\ref{rho*_p_a(t)}), we get an equation for $p_a^\ast$. Once solved and found the value of $p_a^\ast$, we can insert it in Eq.~(\ref{rho*_p_a(t)}) to get $\rho^\ast$.

Clearly, imposing $\rho^\ast=0$ or $\rho^\ast=1$, we get Eqs.~(\ref{tilde_p_a^e})--(\ref{tilde_p_a^p}).

\section{\AB{Results for a population in structured topologies}}
\label{sec:model_SP}

\AB{In this section we explore the effects of considering a social structure governing who interacts with whom within the population. In the previous sections, it has described that, at each time step, a given agent randomly selected a group of $z$ different agents, thus obtaining a variable neighborhood. Here, each agent interacts with a fixed neighborhood of other agents, imposed by the topological structure of a given complex network configuration.}

\begin{figure}[t!]
	\centering
		\includegraphics[width=.85\linewidth]{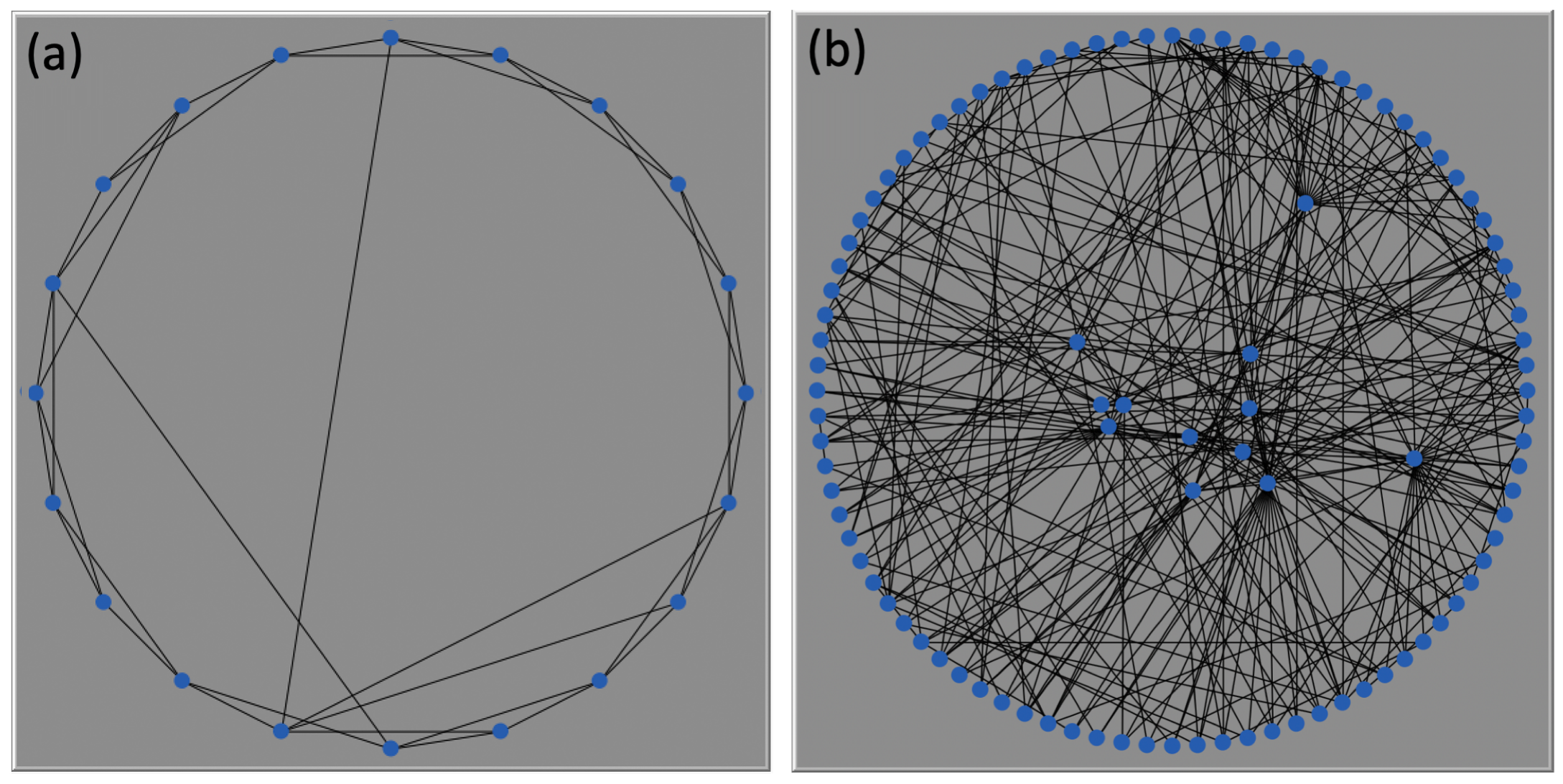}
		\caption{\AP{The complex social structures adopted in this section: (a) the WS4 small-world network and (b) the BA scale-free network. In both cases, a relatively small number of nodes has been chosen for visual convenience.}  
		}
\label{fig:network}
\end{figure}

\AP{The cardinality of the neighborhood of an agent $A_i$, i.e. the number of neighbors of $A_i$, is called degree and is indicated by $k_i$. In general, different agents have different degrees and thus they compare their income with different numbers of other agents. Of course, in a real scenario, any individual generally interacts with a variable subset of their neighborhood. In this work, for sake of simplicity, we assume any agent $A_i$ to interact with all her $k_i$ neighbors at each time step.
The social structure is encoded in the adjacency matrix $\boldsymbol{A}$ of the network, whose element $a_{ij}$ is equal to 1 if agent $A_i$ has agent $A_j$ as a neighbor, and equal to 0 otherwise (loops are here avoided, so $\boldsymbol{A}$ is traceless). We consider undirected networks, thus interactions are symmetric and so is $\boldsymbol{A}$. The degree of node $i$-th is simply computed as $k_i = \sum_j a_{ij}$.}

\AB{In order to explore topologies with different degree distributions, we consider here two of the most celebrated structures in the literature of social networks: the Watts-Strogatz (WS) small-world and the Barabasi-Albert (BA) scale-free, shown in Figures \ref{fig:network}(a) and \ref{fig:network}(b), respectively.} 

\AB{The first one is built by starting from a sequence of $N$ nodes, in which each node is connected to the next and previous two ones. Then, by imposing periodic boundary conditions, each of them has four neighbors. Following the procedure introduced by Watts and Strogatz (1998), we rewire each existent link with a certain probability, typically of order $10^{-2}$, to obtain a Watts-Strogatz small-world network with $\left\langle k\right\rangle = 4$ (WS4), as the one shown in Figure \ref{fig:network}(a). In this kind of network the average distance (the length of the shortest sequence of nodes connecting two given nodes) scales logarithmically with the network size $N$. Loosely speaking, in a small-world network any two nodes are --comparatively to the network size-- close to each other.}

\begin{figure}[t!]
	\centering
		\includegraphics[width=.49\linewidth]{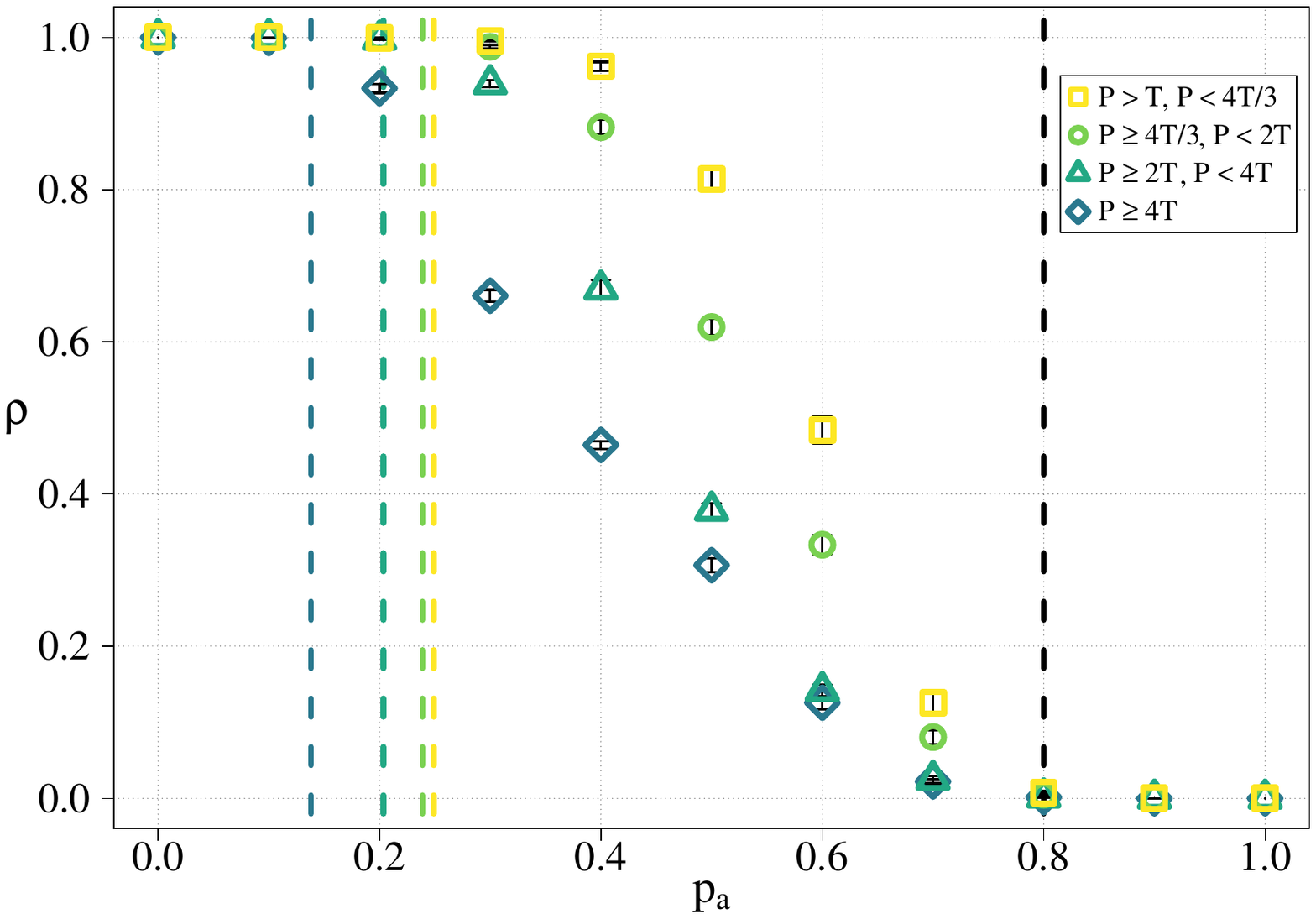}
		\hfill
		\includegraphics[width=.49\linewidth]{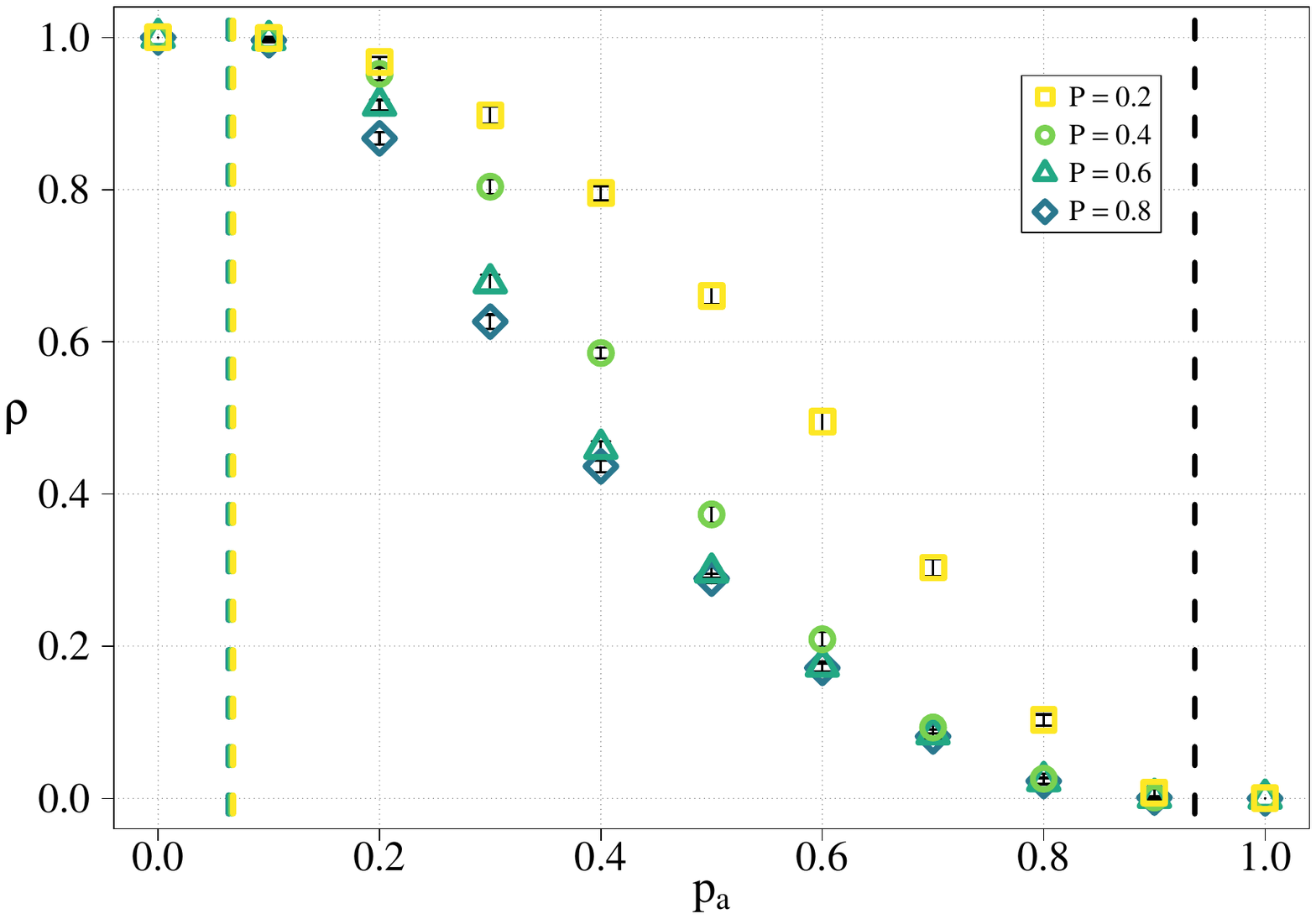} \\
		\includegraphics[width=.49\linewidth]{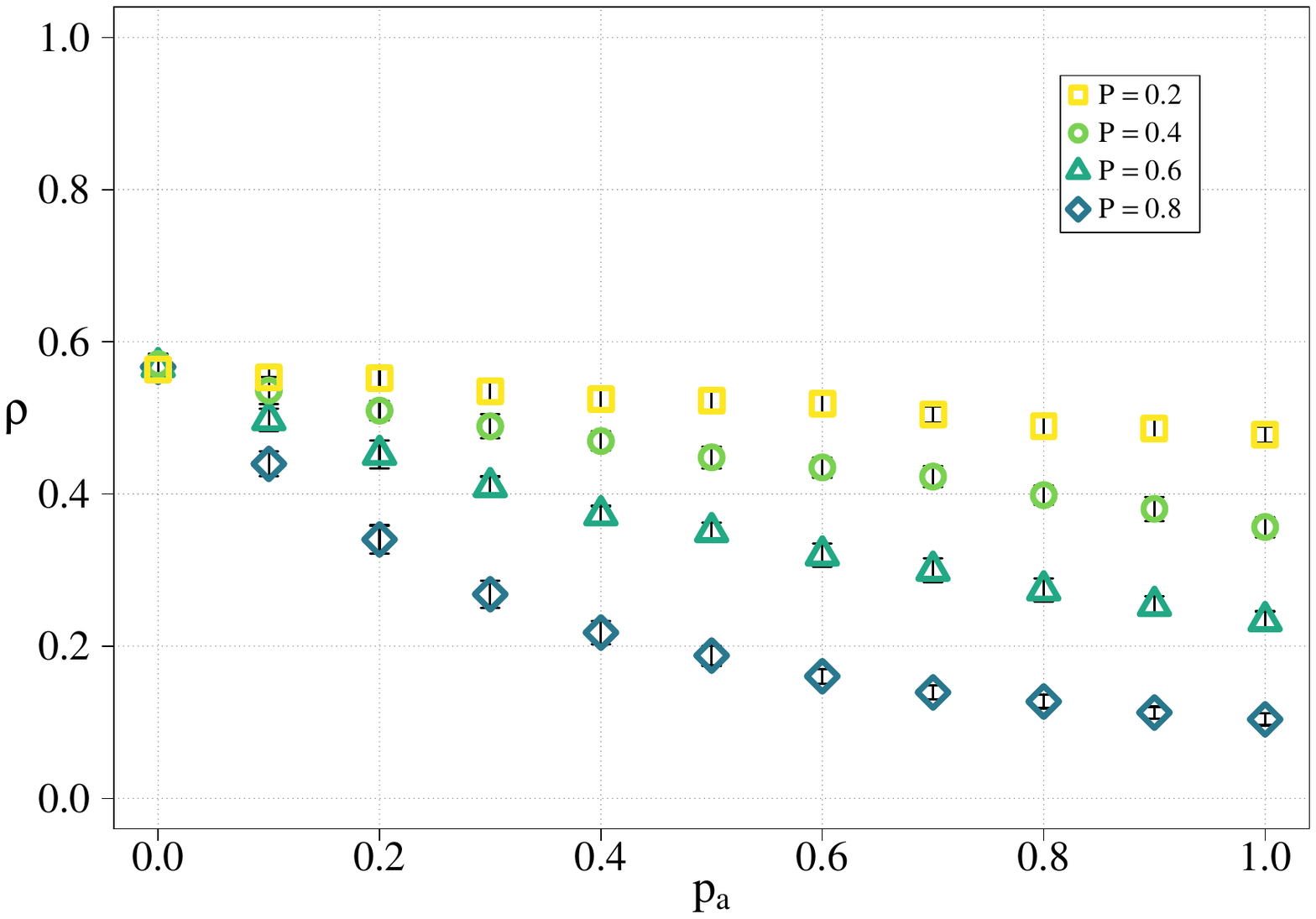}
		\hfill
		\includegraphics[width=.49\linewidth]{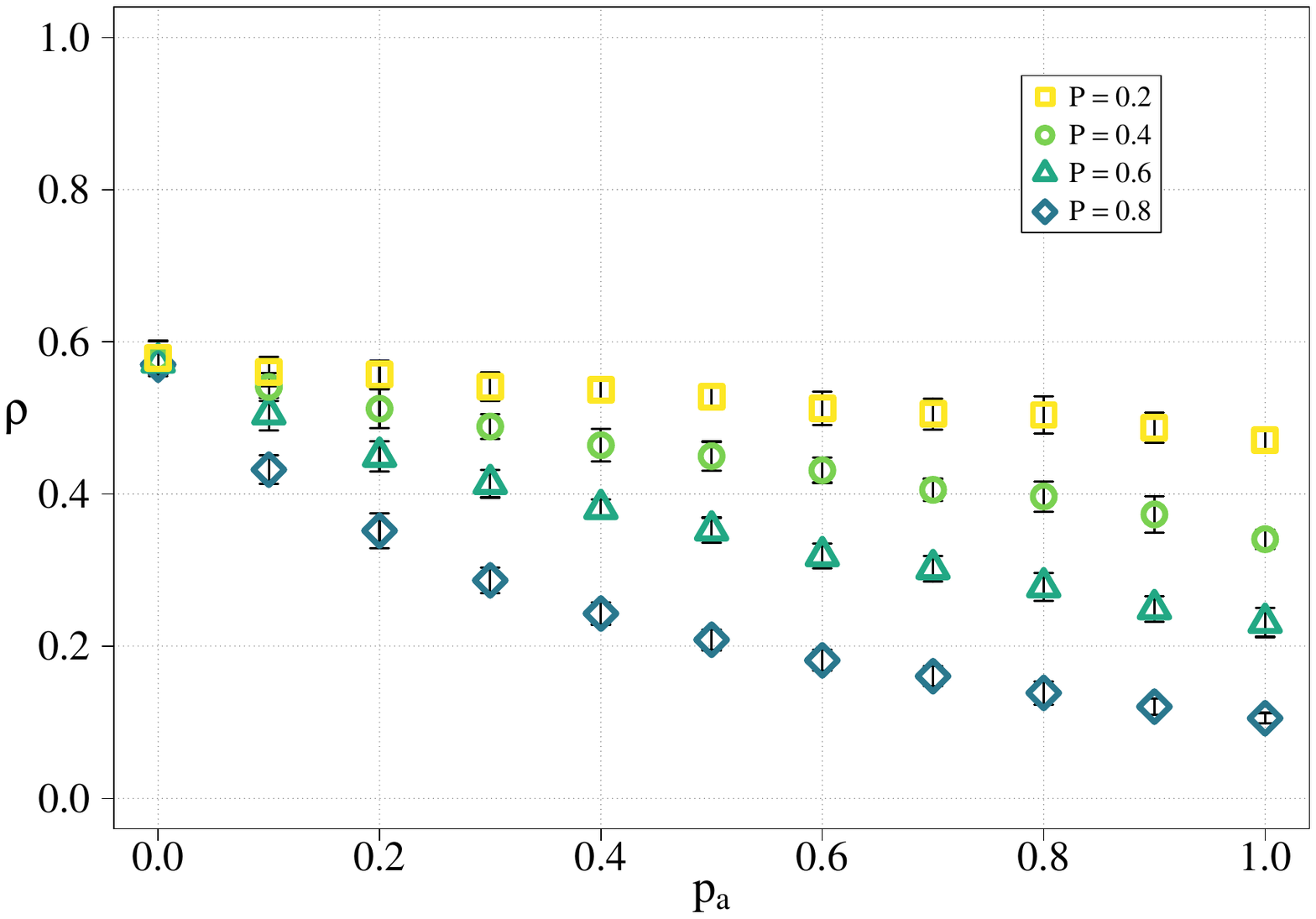}
		
		\caption{Results for a population in structured topologies. Stable equilibrium proportion of tax-evaders $\rho$ versus $p_a$, for the WS4 (left panels) and BA (right panels) networks, both with $N=5000$; $T=0.15$. Points represent the MC's results and bars the respective standard deviations. {\em Top panels.} Results for the delta distribution. Dashed vertical lines represent the predicted critical points, colored for the $P$-dependent tax-payers extinction points and black for the unique tax-evaders extinction point, as computed through Eqs.~(\ref{evaders_ext_th}) and (\ref{payers_ext_th}). The points on the top left panel are got for $P=0.18, 0.25, 0.4, 0.8$. {\em Bottom panels.} Results for a power-law distribution with $\kappa=1.6$, $c_-=1$ and $c_+=100$. No critical points are found in this case.
		}
\label{fig:results_networks}
\end{figure}

\AB{The BA scale-free network of the Barab\'asi-Albert (1999) model, has a power-law degree distribution (analogously to Eq.~(\ref{power-law_dist})) with exponent $3$. This implies that the majority of nodes have few neighbors, while a minority of them --called `hubs`-- possess lots of connections. In Figure \ref{fig:network}(b), hubs have been displaced inside the circular layout to result more visible. Finite networks with $10^3-10^4$ nodes present a finite-size cut-off on the tail of the degree distribution (at $k\approx 50$ for $N=5000$) that makes the degree variance comparable with the mean degree. In particular, we use BA networks with $\left\langle k\right\rangle = 4$ (and variance $\sigma_k \approx 6.2$).} 

\AB{For the sake of computational tractability, we report in Figure~\ref{fig:results_networks} only the results of numerical simulations for WS4 (left panels) and BA (right panels) networks, both with $N=5000$ nodes and for both the income distributions presented in Section~\ref{sec:results}. However, by adopting a Microscopic Markov Chain Approach (MMCA), presented in detail  in Appendix, we are able to provide the analytical estimate of critical points in case of equal incomes. Those critical points will be marked in the plots by vertical solid lines at the corresponding value of $p_a$.}

\AP{Top panels, which refer to the delta income distribution, firstly reveal the goodness of the predictions for the critical points made through the MMCA, as given by Eqs.~(\ref{evaders_ext_th}) and (\ref{payers_ext_th}). It is worth to notice that this is quite unexpected, especially for WS4 networks, where a high level of local clustering could have created non-negligible correlations between the states of clustered neighbors. Moreover, the results look similar for the two networks, although they have very dissimilar properties. Actually, this is observed for both the delta and the power-law income distributions (shown in the bottom panels), suggesting that structural properties of the network play a minor role on the dynamics.} In fact, in the case of equal incomes, the scale-free structure of the BA networks slightly widens the transition region, i.e., makes a bit easier for tax-payers to survive and a bit harder for tax-evaders to die out, in comparison with a regular network of equal mean degree. Overall, we conclude that the homogeneous mean-field approximation of the network, which provides the analytically tractable case of a well-mixed population (see Appendix for details), is robust and reliable.

It is worth noting that, however, this could not be the case if network degree and income of an agent were correlated. This possibility is left to a future work.

\section{Conclusive Remarks}
\label{sec:discussion}

\AB{In this study we propose a new mathematical model of proportional taxation that is able to provide some interesting insights about effective policies for controlling tax evasion while maintaining its analytical tractability.}
	
\AB{Considering the case without any topological structure of interactions among agents, i.e., a well-mixed population, several theoretical predictions have been obtained, with regards to the asymptotic equilibrium states of the fraction of tax-evaders, as a function of the audit probability, and to the critical thresholds above which tax-evaders are expected to disappear. We found that these thresholds depend on the value of the fixed proportional fine in both cases of an \lq\lq egalitarian" and \lq\lq aristocratic" populations, where income is equal for all or power-law distributed, respectively.}

\AB{This finding is interesting because the audit probability is one of the most tunable parameter for the policy-maker, along with tax rate and the amount of fines. Interestingly, in a society where the cost of audits is} \GB{partially (or even almost entirely) sustained by the taxes and the fines,} \AB{implying a variation in time of the audit probability, we obtained that if the amount of the fine is high enough, this is able to set the number of tax-evaders to zero, regardless of the audit probability in an \lq\lq aristocratic" population. Instead, in the \lq\lq egalitarian" case, tax evasion can be completely eradicated only} \GB{if the time-independent baseline of audit probability is high enough.}

\AP{Numerical Monte Carlo simulations confirmed all our analytical predictions, with a very good degree of approximation (specially in the case of delta distribution of income).} 

\AP{Finally, through a Microscopic Markov Chain Approach, we analytically derived the extinction thresholds also for an \lq\lq egalitarian" population with a complex structure. Both a Watts-Strogatz small-world network and a Barabasi-Albert scale-free network have been adopted as realistic models for such a structured population. Simulations revealed a good agreement with the analytical results for the \lq\lq egalitarian" population, thus giving also some insights about the behavior of a structured \lq\lq aristocratic" population with the same complex network structure. Specifically, in case of a power-law distribution of incomes, only a reduction of tax evasion is possible, while its eradication seems impossible, no matter how the fine and the audit probability are set (within the chosen intervals).} 

\AB{In terms of network analysis, the consideration of different topologies for the individual comparison of wealth is relevant in giving a first insight of the behavioral and social nature of tax evasion. The model considers the case in which people can compare their income and decide to imitate successful neighbors with different strategies. The typical case of such an interaction is a tax payer who is persuaded to evade by the positive experiences of her neighbors. A relevant policy implication is to induce people to perceive and underline social respect as a merit good, by means of information, moral suasion, and advertising measures. This calls for educational policies that improve the consciousness in civic engagement.} 

\AP{Even if our results showed a small dependence of the asymptotic behavior of tax-evaders on the chosen network topology (thus confirming the robustness and the reliability of the MMCA approach under the mean-field approximation), structured interactions among citizens could be shown to play a crucial role either when the decision for an agent to change strategy would depend on the cumulated capital more than on the income or \GB{when some kind of correlation between income/capital and structural heterogeneities (e.g., node degree or centrality) would exist}. A forthcoming study will go in this direction, by presenting agent-based simulations for the analytically intractable cases of these more complex dynamics, thus providing policy prescriptions tuned to actual features of contemporary societies.}

\newpage
\section*{APPENDIX}
\label{sec:appendix}

\subsection*{Integrals for  $\rho^*(p_a)$}
\label{sec:app_0}

In the following, the nine integrals $\{{\cal I}_a\}_{a\in\{1,\dots,9\}}$ referred to Eq.\ref{rho*_2}:

\begin{subequations}
\begin{align}
	&{\cal I}_1 = p_a \int_{c_-}^{c_+} dc~p(c) \int_{c_-}^{c_+} dc_1~p(c_1) \int_{2c\frac{1-P}{1-T}-c_1}^{c_+} dc_2~p(c_2) \\
	&{\cal I}_2 = \left(1-p_a\right) \int_{c_-}^{c_+} dc~p(c) \int_{c_-}^{c_+} dc_1~p(c_1) \int_{2c\frac{1}{1-T}-c_1}^{c_+} dc_2~p(c_2) \\
	&{\cal I}_3 = -2\left(1-p_a\right) \int_{c_-}^{c_+} dc~p(c) \int_{c(1-T)}^{c_+} dc_1~p(c_1) \\
	&{\cal I}_4 = -2p_a \int_{c_-}^{c_+} dc~p(c) \int_{c\frac{1-T}{1-P}}^{c_+} dc_1~p(c_1) \\
	&{\cal I}_5 = -2p_a \int_{c_-}^{c_+} dc~p(c) \int_{c\frac{1-P}{1-T}}^{c_+} dc_1~p(c_1) \\
	&{\cal I}_6 = -2\left(1-p_a\right) \int_{c_-}^{c_+} dc~p(c) \int_{c\frac{1}{1-T}}^{c_+} dc_1~p(c_1) \\
	&{\cal I}_7 = \left(1-p_a\right)^2 \int_{c_-}^{c_+} dc~p(c) \int_{c_-}^{c_+} dc_1~p(c_1) \int_{2c(1-T)-c_1}^{c_+} dc_2~p(c_2) \\
	&{\cal I}_8 = 2p_a\left(1-p_a\right) \int_{c_-}^{c_+} dc~p(c) \int_{c_-}^{c_+} dc_1~p(c_1) \int_{\frac{2c(1-T)-c_1}{1-P}}^{c_+} dc_2~p(c_2) \\
	&{\cal I}_9 = p_a^2 \left(1-p_a\right) \int_{c_-}^{c_+} dc~p(c) \int_{c_-}^{c_+} dc_1~p(c_1) \int_{2c\frac{1-T}{1-P}-c_1}^{c_+} dc_2~p(c_2)
	\label{I_a_2}
\end{align}
\end{subequations}

\subsection*{Critical Points for $\boldsymbol {z=1}$}
\label{sec:app_A}

In this section we derive the critical point $p_a^{(e)}=p_a^{(p)}$, marking the first-order phase transition existent for $z=1$, for the power-law income distribution defined in Eq.~(\ref{power-law_dist}). We recall the transition for the case of equal incomes to be at $1/2$, as estimated in Section 3.

Eq.~(\ref{p_a^e}), evaluated for $z=1$, gives
\begin{equation}
	p_a^{(e)}\left(1,p(c)\right) = \frac{\displaystyle\int_{c_-}^{c_+} dc~p(c) \int_{c(1-T)}^{\frac{c}{1-T}} dc_1~p(c_1)}{\displaystyle\int_{c_-}^{c_+} dc~p(c) \left[\int_{c(1-T)}^{c\frac{1-T}{1-P}} dc_1~p(c_1) + \int_{c\frac{1-P}{1-T}}^{\frac{c}{1-T}} dc_1~p(c_1)\right]}
\label{p_a^e_z=1}
\end{equation}
With some algebra, one gets the critical value reported below. \\

\subsubsection*{Power-law Distribution.}
\begin{equation}
	\notag p_a^{(e)}\left(1,p_\kappa(c)\right) = \left\{1 + \frac{\displaystyle\Theta\left[\frac{1-P}{1-T}c_+-c_-\right]\left(A_1 - A_2\right) + \frac{\left({c_+^{1-\kappa}-c_-^{1-\kappa}}\right)^2}{1-\kappa}}{\displaystyle\Theta\left[\left(1-T\right)c_+-c_-\right]\left(B_1 - B_2\right) + \frac{\left({c_+^{1-\kappa}-c_-^{1-\kappa}}\right)^2}{1-\kappa}}\right\}^{-1}
\label{p_a^e_z=1_power-law}
\end{equation}
where
\begin{subequations}
\begin{align}
	\notag &A_1 \equiv \left({\frac{1-T}{1-P}}\right)^{1-\kappa}\frac{\left({\frac{1-P}{1-T}}\right)^{2\left(1-\kappa\right)}c_+^{2\left(1-\kappa\right)}-c_-^{2\left(1-\kappa\right)}}{2\left(1-\kappa\right)} + c_+^{2\left(1-\kappa\right)}\frac{1-\left({\frac{1-P}{1-T}}\right)^{1-\kappa}}{1-\kappa} - c_+^{1-\kappa}\frac{c_+^{1-\kappa}-c_-^{1-\kappa}}{1-\kappa} \\
	\notag &A_2 \equiv \left({\frac{1-P}{1-T}}\right)^{1-\kappa}\frac{c_+^{2\left(1-\kappa\right)}-\left({\frac{1-T}{1-P}}\right)^{2\left(1-\kappa\right)}c_-^{2\left(1-\kappa\right)}}{2\left(1-\kappa\right)} + c_-^{2\left(1-\kappa\right)}\frac{\left({\frac{1-T}{1-P}}\right)^{1-\kappa}-1}{1-\kappa} - c_-^{1-\kappa}\frac{c_+^{1-\kappa}-c_-^{1-\kappa}}{1-\kappa} \\
	\notag &B_1 \equiv \left({\frac{1}{1-T}}\right)^{1-\kappa}\frac{\left(1-T\right)^{2\left(1-\kappa\right)}c_+^{2\left(1-\kappa\right)}-c_-^{2\left(1-\kappa\right)}}{2\left(1-\kappa\right)} + c_+^{2\left(1-\kappa\right)}\frac{1-\left(1-T\right)^{1-\kappa}}{1-\kappa} - c_+^{1-\kappa}\frac{c_+^{1-\kappa}-c_-^{1-\kappa}}{1-\kappa} \\
	\notag &B_2 \equiv \left(1-T\right)^{1-\kappa}\frac{c_+^{2\left(1-\kappa\right)}-\left({\frac{1}{1-T}}\right)^{2\left(1-\kappa\right)}c_-^{2\left(1-\kappa\right)}}{2\left(1-\kappa\right)} + c_-^{2\left(1-\kappa\right)}\frac{\left(\frac{1}{1-T}\right)^{1-\kappa}-1}{1-\kappa} - c_-^{1-\kappa}\frac{c_+^{1-\kappa}-c_-^{1-\kappa}}{1-\kappa}
\label{pieces_power-law}
\end{align}
\end{subequations}

\subsection*{Microscopic Markov Chain Approach for Equal Incomes}
\label{sec:app_B}

We here adapt the Microscopic Markov Chain Approach (MMCA) developed by G\'omez et al. (2010) to model spreading diseases, to our behavioral economics problem. The MMCA gives a microscopic description able to account for the actual structure of interactions among the agents, that is, for who interacts with whom. In the following, we consider a population whose individuals get all the same income.

In the herein proposed adaptation, the MMCA describes the time evolution of the probability $p_i(t)$ that the $i$-th agent is a tax-evader at time $t$. The state of the system is thus defined by the vector of probabilities $\boldsymbol{p}(t)=\left(p_1(t),\dots,p_N(t)\right)$ or, as a coarse-grained, global variable, by the expected fraction of tax-evaders
\begin{equation}
	\rho(t) = \frac1N \sum_{i=1}^N p_i(t)
\label{rho_MMCA}
\end{equation}

To follow the evolution of the population we need a dynamic equation for the single $p_i(t)$, $\forall i\in\{1,\dots,N\}$. All the considerations about the strategy update made before for a well-mixed population, hold also here, with just few modifications accounting for the structure.

The equation for $p_i(t)$ has an inflow and an outflow term of the same form of those given in Section 3 for a delta income distribution, but expressed through the entries of $\boldsymbol{p}(t)$ and $\boldsymbol{A}$. It reads
\begin{equation}
	\frac{dp_i(t)}{dt} = -~p_i(t) p_a\left[1-\prod_{j=1}^N a_{ij}p_j(t)\right] + \left[1-p_i(t)\right] {\cal Q}^{p\rightarrow e}_{(i)}\left(\boldsymbol{p}(t);p_a,P,T\right)
\label{dyn_eq_p_i}
\end{equation}
where ${\cal Q}^{p\rightarrow e}_{(i)}\left(\boldsymbol{p}(t);p_a,P,T\right)$ is the probability for agent $A_i$ to become a tax-evader when it is currently a tax-payer.

The product in the outflow term reveals the basic approximation made in the MMCA, that is, that the states of the agents are assumed to be uncorrelated variables. So, for example, the probability for a set of agents to be all tax-evaders is computed as the product of the probabilities for each of them to be a tax-evader. The assumed local states' independence is the unique source of error in the MMCA's predictions.

\begin{figure}[tb!]
	\centering
		\includegraphics[width=.74\linewidth]{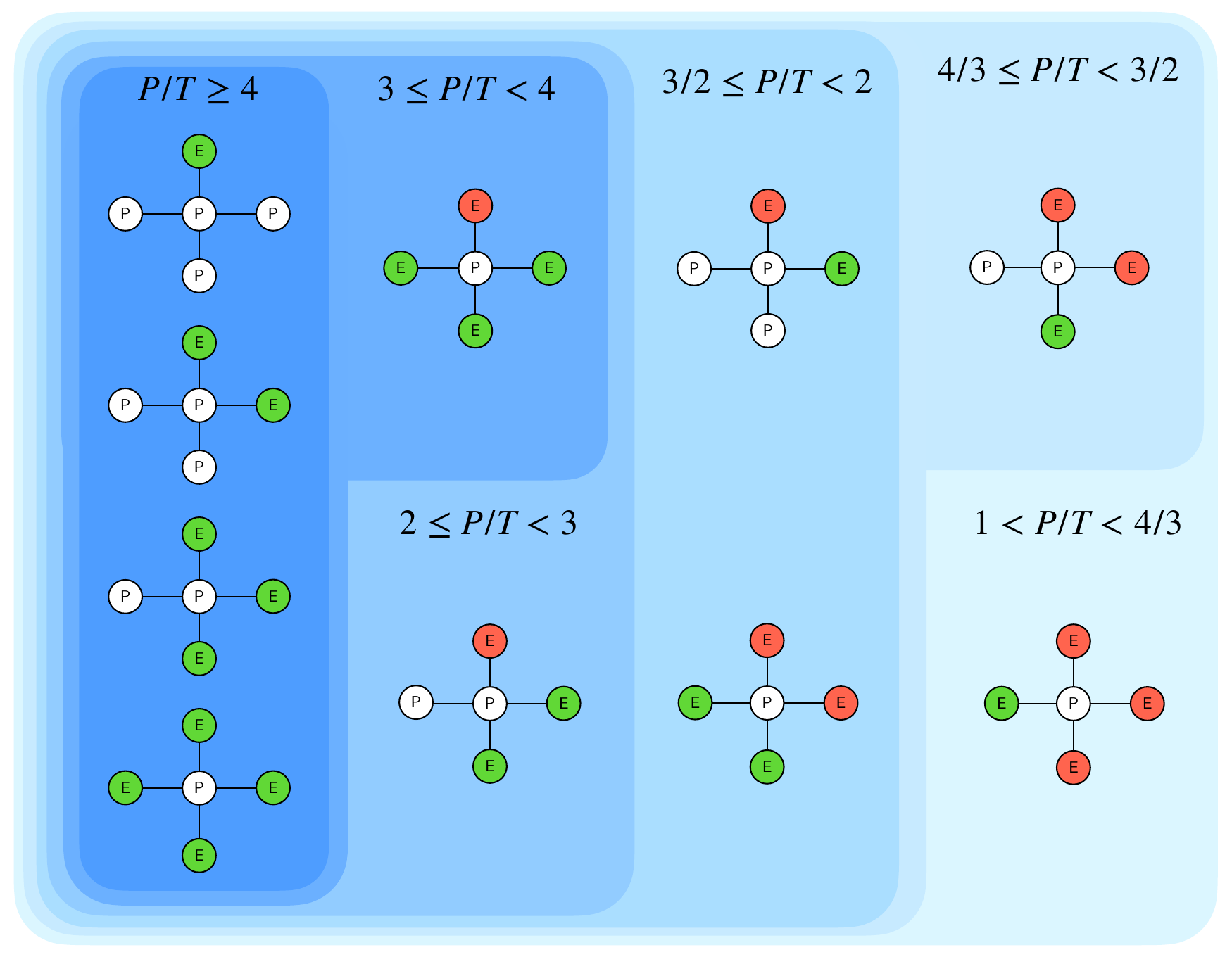}
		
		\caption{All the possible local configurations contributing to ${\cal Q}^{p\rightarrow e}$ for a tax-payer (P) having four neighbors, depending on the value of the ratio $P/T$. Tax-evaders (E) can be \lq\lq lucky\rq\rq (green) or \lq\lq unlucky\rq\rq(red).
		}
\label{fig:conf_z=4}
\end{figure}

The expression of ${\cal Q}^{p\rightarrow e}_{(i)}$ is node-dependent, for it depends on the degree $k_i$ of node $i$. Its general expression takes the form
\begin{align}
	\notag {\cal Q}^{p\rightarrow e}_{(i)}&\left(\boldsymbol{p}(t);p_a,P,T\right) = \sum_{n_e = 1}^{k_i} \sum_{n_e^- = 0}^{\left\lfloor{n_eT}/P\right\rfloor} \left(1-p_a\right)^{n_e-n_e^-}p_a^{n_e^-} \times \\
	& \times \frac1{(k_i-n_e)!(n_e-n_e^-)!(n_e^-)!} \underset{j_1,\dots ,j_{k_i}}{\sum\nolimits^*\;}a_{ij_1}\cdots a_{ij_{k_i}}p_{j_1}(t)\cdots p_{j_{n_e}}(t)[1-p_{j_{{n_e}+1}}(t)]\cdots[1-p_{j_{k_i}}(t)]
\label{Q_(i)}
\end{align}
where the constant term in front of the last series of sums is a normalization factor to avoid multiple counting of the same configuration, while the notation $\sum\nolimits^*$ is a shortcut for the constraint $j_h\neq j_l$, $\forall \{h,l\}\subseteq\{1,\dots,k_i\}$. As an example, for $k_i = 4$, the contributing terms to Eq.~(\ref{Q_(i)}) are those represented in Figure~\ref{fig:conf_z=4} taking $i$ as the central, tax-payer node.

Once initialized the vector $\boldsymbol{p}(0)$ respecting the constraint imposed by $\rho(0)$, the Markov chain is ready to be run, till convergence to a stationary configuration.

The microscopic description given by the MMCA -- or by other approaches of comparable detailedness -- is very useful for the calculation of the critical points of the dynamics for any topology. This depth of detail, however, comes with the drawback of a high computational cost. In our model, in particular, what greatly enhances this cost is the many-agent interaction defining the strategy update. Indeed, looking at Eq.~(\ref{Q_(i)}), the sum over the indices $\{j_1,\dots,j_{k_i}\}$ implies the computation of ${k_i}!$ terms for agent $A_i$. Indicating with $\left\langle x\right\rangle$ the network average of quantity $x$, it means computing $N\left\langle k!\right\rangle$ terms, at each time step. This is generally a very large number when considering large populations ($N\approx10^3$, at least), particularly when dealing with (even slightly) heterogeneous topologies mimicking real social interaction patterns. Additionally, the fact that the number and the form of the terms in ${\cal Q}^{p\rightarrow e}_{(i)}$ depend on the degree of the node, makes extremely hard to evaluate it even for small heterogeneities. Therefore, the MMCA is generally ineffective to get any non-ab-initio prediction in a feasible amount of time, except for very homogeneous structures with low average degree. By the way, regarding the latter, no significant differences are expected in comparison with the results obtained for a related well-mixed population (taking $z$ equal to the average degree of the network), as the critical points are predicted to be the same (as shown below). More generally, the critical points for a structured population are expected to be close to those of the respective unstructured one, whenever the degree variance of the network is small compared to its squared degree average (that is, $\left\langle k^2\right\rangle \approx {\left\langle k\right\rangle}^2$).

It is worth to note that Eq.~(\ref{dyn_eq_rho}), holding for a well-mixed population, can be recovered as the homogeneous mean-field approximation of Eq.~(\ref{dyn_eq_p_i}), by taking all the nodes as equivalent and neglecting the states' fluctuations, i.e., $k_i=z$ (homogeneity) and $p_i = \rho$, $\forall i \in \{1,\dots ,N\}$. Indeed, making this approximation, the structure tacitly disappears and the critical thresholds for a well-mixed population are recovered.

That said, in the following we limit ourselves to derive the critical thresholds, for any topology. We consider the more general case of time-dependent $p_a$, while the time-independent case is recovered simply taking $\beta = 0$ in Eq.~(\ref{p_a(t)}). \\

\subsubsection*{\bf Tax-Evaders Extinction Thresholds}
\label{par:evaders_ext}

We derive here the threshold values for $p_a$ and $T$ above which tax-evaders are predicted to be extinct.

We first linearize Eq.~(\ref{dyn_eq_p_i}) taking $p_i = \epsilon_i \ll 1$, $\forall i\in\{1,\dots,N\}$. Among all the local configurations contributing to ${\cal Q}^{p\rightarrow e}_{(i)}$, only those of order $\epsilon$ are non-negligible. Thus, with few algebra, Eq.~(\ref{dyn_eq_p_i}) becomes
\begin{equation}
	\frac{d\epsilon_i(t)}{dt} = - p_a\epsilon_i(t) + (1-p_a) \sum_{j=1}^N a_{ij}\epsilon_j(t) = \sum_{j=1}^N \left[(1-p_a)a_{ij}-p_a\delta_{ij}\right]\epsilon_j(t)
	\label{dyn_eq_p_i_lin}
\end{equation}
where $ \delta_{ij}$ is the $(i,j)$ entry of the $N\times N$ identity matrix $\boldsymbol{ \mathbb{I}}_{N\times N}$. Imposing the stationarity condition, we find an eigenvalue equation. Additionally, excluding the trivial case $p_a=1$, for which $\rho=0$ is the only stable equilibrium solution, dividing by $1-p_a$, one gets
\begin{equation}
	\sum_{j=1}^N \left[a_{ij}-\frac{p_a}{1-p_a}\delta_{ij}\right]\epsilon_j = 0
	\label{eigen_lin}
\end{equation}
from which it follows $p_a/(1-p_a)$ to be an eigenvalue of the adjacency matrix $\boldsymbol A$. In particular, being $p_a/(1-p_a)$ an increasing function of $p_a$, we estimate the tax-evaders extinction threshold $p_a^{(e)}$, above which no tax-evaders are expected to survive at the equilibrium, as related to the largest eigenvalue $\Lambda_{max}(\boldsymbol A)$ of the adjacency matrix of the network, that is
\begin{equation}
	p_a^{(e)} = \frac{\Lambda_{max}(\boldsymbol A)}{\Lambda_{max}(\boldsymbol A)+1}
	\label{evaders_ext_th}
\end{equation}
For a $z$-regular network, $\Lambda_{max}(\boldsymbol A) = z$, and hence $p_a^{(e)} = z/(z+1)$, as found for a well-mixed population.

Since, from Eq.~(\ref{p_a(t)}), $p_a$ reduces to a constant in the limit $\rho = \epsilon \ll 1$, Eq.~(\ref{evaders_ext_th}) holds also when $p_a$ depends on time in that way.

Finally the same expression for $T^{(e)}$ as given by Eq.~(\ref{T_e}) follows taking $\rho=0$ in Eq.~(\ref{p_a(t)}). \\

\subsubsection*{\bf Tax-Payers Extinction Thresholds}
\label{par:payers_ext}

We derive here the threshold values for $p_a$ and $P$ below which tax-payers are predicted to be extinct.

As before, we start by linearizing Eq.~(\ref{dyn_eq_p_i}), but now around $\rho=1$, taking $p_i = 1-\epsilon_i$, $\forall i\in\{1,\dots,N\}$, with $\epsilon_i \ll 1$. Since $1-p_i$ is of order $\epsilon$, the non-negligible contributions to ${\cal Q}^{p\rightarrow e}_{(i)}$ come from only those configurations with no tax-payers, i.e., $n_e = k_i$. Thus, Eq.~(\ref{dyn_eq_p_i}) becomes
\begin{align}
	\notag \frac{d\epsilon_i(t)}{dt} &= p_a\sum_{j=1}^N a_{ij}\epsilon_j(t) - \epsilon_i(t)\left[{\cal Q}^{p\rightarrow e}_{(i)}\left(\boldsymbol{p}(t);p_a,P,T\right)\right]_{n_e = k_i} \\
	\notag &= p_a\sum_{j=1}^N a_{ij}\epsilon_j(t) - \epsilon_i(t)\sum_{n_e^- = 0}^{\left\lfloor{k_iT}/P\right\rfloor}\frac{k_i!}{(k_i-n_e^-)!(n_e^-)!}\left(1-p_a\right)^{k_i-n_e^-}p_a^{n_e^-} \\
	&= \sum_{j=1}^N \left[p_aa_{ij} - \left(\sum_{n_e^- = 0}^{\left\lfloor{k_iT}/P\right\rfloor}\frac{k_i!}{(k_i-n_e^-)!(n_e^-)!}\left(1-p_a\right)^{k_i-n_e^-}p_a^{n_e^-}\right)\delta_{ij}\right]\epsilon_j(t)
	\label{dyn_eq_p_i_lin_2}
\end{align}
Imposing the stationarity condition we get an eigenvalue equation. In particular, rearranging it excluding the trivial case $p_a=0$, for which $\rho=1$ is the only stable equilibrium solution, and observing the term in round brackets to be strictly positive, we can write
\begin{equation}
	\sum_{j=1}^N \left[\left(\sum_{n_e^- = 0}^{\left\lfloor{k_iT}/P\right\rfloor}\frac{k_i!}{(k_i-n_e^-)!(n_e^-)!}\left(1-p_a\right)^{k_i-n_e^-}p_a^{n_e^-}\right)^{-1}a_{ij} - p_a^{-1}\delta_{ij}\right]\epsilon_j = 0
	\label{payers_ext_eigen}
\end{equation}
which form depends on the ratio $T/P$. The threshold value is then implicitly defined as an eigenvalue of the matrix $\boldsymbol{A^\prime}(p_a)$, of elements
\begin{equation}
	a^\prime_{ij}(p_a) = \left(\sum_{n_e^- = 0}^{\left\lfloor{k_iT}/P\right\rfloor}\frac{k_i!}{(k_i-n_e^-)!(n_e^-)!}\left(1-p_a\right)^{k_i-n_e^-}p_a^{n_e^-}\right)^{-1}a_{ij}
	\label{A^prime}
\end{equation}
Now, for any connected and undirected network, $\boldsymbol{A}$ is an irreducible, non-negative matrix (see Meyer (2000)). The Perron-Frobenius theorem thus ensures that $\Lambda_{max}(\boldsymbol A)>0$ and, moreover, that $\underset{i}{\textup {min}}~k_i\leq\Lambda_{max}(\boldsymbol A)\leq\underset{i}{\textup {max}}~k_i$. Since the network is connected, it follows $\Lambda_{max}(\boldsymbol A)\geq1$. Furthermore, being the term $\left(\cdots\right)^{-1}$ in Eq.~(\ref{A^prime}) a continuous, strictly increasing function of $p_a$, taking value $1$ at $p_a=0$, it follows $a^\prime_{ij}(p_a)\geq a_{ij}$, $\forall \{i,j\}$. Then, being $\boldsymbol{A^\prime}(p_a)$ irreducible as well, from the Wielandt's theorem (Meyer (2000)) it follows $\Lambda_{max}\left(\boldsymbol A^\prime\left(p_a\right)\right)$ to be a strictly increasing function of $p_a$ and hence $\Lambda_{max}\left(\boldsymbol A^\prime\left(p_a\right)\right) \geq \Lambda_{max}(\boldsymbol A) \geq 1$, with the first inequality being strict for any $p_a>0$. At this point, observing that $p_a^{-1}$ is a decreasing function of $p_a$, going continuously from $+\infty$ at $p_a = 0$ to $1$ at $p_a = 1$, from what just proved, it follows that it does exist $p_a = p_a^{(p)}$ at which $p_a^{-1}$ intersects $\Lambda_{max}\left(\boldsymbol A^\prime\left(p_a\right)\right)$, that is
\begin{equation}
	p_a^{(p)} = \frac{1}{\Lambda_{max}\left(\boldsymbol A^\prime\left(p_a^{(p)}\right)\right)}
	\label{payers_ext_th}
\end{equation}
Eq.~(\ref{payers_ext_th}) provides implicitly the smallest value of $p_a$ below (above) which tax-payers are expected to be extinct (to survive).

To give an example of how Eq.~(\ref{payers_ext_th}) works, let us briefly consider the case of a $z$-regular network (identical thresholds follow for a well-mixed population) and $T/P \leq z^{-1}$, for which the only contributing term is that with $n_e^- = 0$. We thus get ${a^\prime_{ij}(p_a)} = (1-p_a)^{-z}a_{ij}$, and $\Lambda_{max}(\boldsymbol A^\prime(p_a)) = z(1-p_a)^{-z}$. Inserting this in Eq.~(\ref{payers_ext_th}), we get $p_a^{(p)}$ as the solution of $zp_a-(1-p_a)^{z}=0$. The other tax-payers extinction thresholds can be computed in the same manner.

At last, taking $\rho=1$ in Eq.~(\ref{p_a(t)}), we find the expression given in Eq.~(\ref{P_p}) for $P^{(p)}$.

As discussed before, according to Eq.~(\ref{p_a(t)}), the critical properties of the system for that time-dependent form of $p_a$, are unchanged.

\newpage
\section*{References}

Allingham, M. G. and Sandmo, A. (1972). Income tax evasion: a theoretical analysis. \textit{Journal of Public Economics}, vol.1(3-4), pp.323-338.

Alstadsæter, A., Jacob, M., \& Michaely, R. (2017). Do dividend taxes affect corporate investment?. \textit{Journal of Public Economics}, vol.151, pp. 74-83.

Alm, J. (2012). Measuring, explaining, and controlling tax evasion: lessons from theory, experiments, and field studies. \textit{International tax and public finance}, vol. 19(1), pp. 54-77.

Alm, J., Bloomquist, K. M., \& McKee, M. (2017). When you know your neighbour pays taxes: Information, peer effects and tax compliance. \textit{Fiscal Studies}, 38(4), 587-613.

Andreoni, J., Erard, B., and Feinstein, J. (1998). Tax compliance. \textit{Journal of Economic Literature}, vol.36(2), pp.818-860.

Barabasi, A.L. and Albert, R., (1999). Emergence of scaling in random networks. \textit{Science}, 286(5439), pp.509-512.

Bazart, C., Bonein, A., Hokamp, S., \& Seibold, G. (2016). Behavioural economics and tax evasion: calibrating an agent-based econophysics model with experimental tax compliance data. \textit{Journal of Tax Administration}, vol. 2(1), pp. 126-144.

Bertotti, M. L., \& Modanese, G. (2014). Mathematical models for socio-economic problems. In \textit{Mathematical Models and Methods for Planet Earth} (pp. 123-134). Springer.	

Bertotti, M. L., \& Modanese, G. (2016). Microscopic models for the study of taxpayer audit effects. \textit{International Journal of Modern Physics C}, 27(09), 1650100.

Biondo, A. E. (2019). Order book modeling and financial stability. \textit{Journal of Economic Interaction and Coordination}, 14(3), pp.469-489.

Biondo, A. E., Pluchino, A., Rapisarda, A., \& Helbing, D. (2013). Reducing financial avalanches by random investments. \textit{Physical Review E}, 88(6), 062814.

Bloomquist, K. M. (2006). A comparison of agent-based models of income tax evasion. \textit{Social Science Computer Review}, vol. 24, pp.411-425.

Callen, E., \& Shapero, D. (1974). A theory of social imitation. \textit{Physics Today}, vol. 27(7), pp. 23.

Clotfelter, C. T. (1983). Tax evasion and tax rates: An analysis of individual returns. \textit{The review of economics and statistics}, pp. 363-373.

Crane, S. E., \& Nourzad, F. (1987). On the treatment of income tax rates in empirical analysis of tax evasion. \textit{Kyklos}, vol. 40(3), pp. 338-348.

Dalamagas, B. (2011). A dynamic approach to tax evasion. \textit{Public Finance Review}, vol. 39(2), pp. 309-326.

Dubin, J. A. (2007). Criminal investigation enforcement activities and taxpayer noncompliance. \textit{Public Finance Review}, 35:4. 

Elsenbroich, C., \& Gilbert, N. (2014). \textit{Modelling Norms}. Springer, Dordrecht

Epstein, R. A. (1993). Altruism: Universal and Selective. \textit{Social Service Review}, vol. 67(3), pp. 388-405.

Feld, L. P., \& Frey, B. S. (2002). Trust breeds trust: How taxpayers are treated. \textit{Economics of governance}, vol. 3(2), pp. 87-99.

Gómez, S., Arenas, A., Borge-Holthoefer, J., Meloni, S., \& Moreno, Y. (2010). Discrete-time Markov chain approach to contact-based disease spreading in complex networks. \textit{EPL (Europhysics Letters)}, 89(3), 38009. 

R. Hardin, (1995) \textit{One for All: The Logic of Group Conflict}, Princeton University Press, Princeton, NJ.

Heckathorn, D. D. (1996). The dynamics and dilemmas of collective action. \textit{American sociological review}, pp. 250-277.

Hokamp, S., \& Pickhardt, M. (2010). Income tax evasion in a society of heterogeneous agents–Evidence from an agent-based model. \textit{International Economic Journal}, vol. 24(4), pp. 541-553.

Hokamp, S. (2013). Income tax evasion and public goods provision–theoretical aspects and agent-based simulations (Doctoral dissertation, Ph. D. thesis, Brandenburg University of Technology Cottbus).

IRS - Internal Revenue Service Tax Gap Estimates for Tax Years 2008–2010, \\https://www.irs.gov/PUP/newsroom/ tax gap estimates for 2008 through 2010.pdf (accessed November 23, 2016)

Kirchler, E. (2007). \textit{The Economic Psychology of Tax Behaviour}. Cambridge University Press. Cambridge Books Online.

McDonald, R. I., \& Crandall, C. S. (2015). Social norms and social influence. \textit{Current Opinion in Behavioral Sciences}, vol. 3, pp. 147-151.

McGee, R. (2012). \textit{The Ethics of Tax Evasion: Perspectives in Theory and Practice}. SpringerLink : B{\''u}cher. Springer New York. 

Meyer, C. D., (2000) {\em Matrix Analysis and Applied Linear Algebra} Siam. 

Nicolaides, P. (2014). Tax compliance social norms and institutional quality: an evolutionary theory of public good provision. \textit{Directorate General Taxation and Customs Union, European Commission} (No. 46).

Panades, J. (2004). Tax evasion and relative tax contribution. \textit{Public Finance Review}, vol. 32(2), pp. 183-195.

Pickhardt, M. and Prinz, A. (2012). \textit{Tax Evasion and the Shadow Economy}. Edward Elgar Pub. 

Pickhardt, M., \& Prinz, A. (2014). Behavioral dynamics of tax evasion–A survey. \textit{Journal of Economic Psychology}, vol. 40, pp. 1-19.

Pluchino, A., Rapisarda, A., \& Garofalo, C. (2010). The Peter principle revisited: A computational study. \textit{Physica A: Statistical Mechanics and its Applications}, 389(3), pp.467-472.

Pluchino, A., Biondo, A. E., \& Rapisarda, A. (2018). Talent versus luck: The role of randomness in success and failure. \textit{Advances in Complex Systems}, 21(03n04), 1850014.

Poterba, J. M. (1987). How burdensome are capital gains taxes?: Evidence from the United States. \textit{Journal of Public Economics}, vol. 33(2), pp. 157-172.

Rapoport, A. (1974). Prisoner's dilemma – recollections and observations, in \textit{Game Theory as a Theory of Conflict Resolution}, edited by A. Rapoport (Reidel, Dordrecht, The Netherlands, 1974), pp. 17–34

Riahi-Belkaoui, A. (2004). Relationship between tax compliance internationally and selected determinants of tax morale. \textit{Journal of International Accounting, Auditing and Taxation}, vol. 13(2) pp.135-143. 

Shu, L. L., Mazar, N., Gino, F., Ariely, D., and Bazerman, M. H. (2012). Signing at the beginning makes ethics salient and decreases dishonest self-reports in comparison to signing at the end. \textit{Proceedings of the National Academy of Sciences}, 109(38):15197–15200.

Slemrod, J., \& Yitzhaki, S. (2002). Tax avoidance, evasion, and administration. In \textit{Handbook of public economics} (Vol. 3, pp. 1423-1470). Elsevier.

Slemrod, J. (2007). Cheating ourselves: The economics of tax evasion. \textit{The Journal of Economic Perspectives}, vol. 21(1), pp. 25-48.

Squazzoni, F., Jager, W., and Edmonds, B. (2014). Social simulation in the social sciences: A brief overview. \textit{Social Science Computer Review}, vol. 32(3), pp. 279-294.

Stevens, J.B. (2018), \textit{The Economics Of Collective Choice}, Routledge, NY.

Torgler, B. (2002). Speaking to theorists and searching for facts: Tax morale and tax compliance in experiments. \textit{Journal of Economic Surveys}, vol. 16(5), pp.657-683.

Torgler,B.(2007). \textit{Tax Compliance and Tax Morale: A Theoretical and Empirical Analysis}. Edward Elgar Pub.

Torgler, B. and Schaltegger, C. A. (2005). Tax morale and fiscal policy. \textit{Centre for Research in Economics, Management and the Arts Working Paper}, 30.

Torgler, B., \& Schneider, F. (2009). The impact of tax morale and institutional quality on the shadow economy. \textit{Journal of Economic Psychology}, vol. 30(2), pp. 228-245.

Vale, R. (2015). A model for tax evasion with some realistic properties. Available at \textit{SSRN} 2601214.

Voss, T. (2001) Game-theoretical perspectives on the emergence of social norms, in \textit{Social Norms}, edited by M.K.D. Hechter, Opp (Russell Sage Foundation, New York, 2001), pp. 105-136.

Yitzhaki, S. (1987). On the excess burden of tax evasion. \textit{Public Finance Quarterly}, vol.15(2), pp.123-137.

Watts, D.J. and Strogatz, S.H., (1998). Collective dynamics of 'small-world'networks. \textit{Nature}, 393(6684), pp.440-442.

Weidlich, W. (1991). Physics and social science—the approach of synergetics. \textit{Physics reports}, 204(1), pp.1-163.

\end{document}